\documentclass[letterpaper,titlepage,11pt]{article}
\usepackage{hyperref}
\usepackage{amssymb,amsmath,amsfonts}
\usepackage{epsfig}

\def\clock{{\count0=\time
           \divide\count0 60
           \ifnum\count0<10 0\fi\the\count0
           \multiply\count0 -60 \advance\count0 \time
           :\ifnum\count0<10 0\fi \the\count0
         }}
\newcommand{\timestamp}{{\small\vbox{\hbox{\tt\jobname.tex}
\hbox{\the\day/\the\month/\the\year, \clock}}}}

\newcommand{\CO}{\mathcal{O}}
\newcommand{\CN}{\mathcal{N}}

\newcommand{\CM}{\mathcal{M}}

\newcommand{\Z}{\mathbb{Z}}

\newcommand{\R}{\mathbb{R}}

\newcommand{\nn}{\nonumber}

\newcommand{\spa}{\ , \ \ }

\newcommand{\ds}{\displaystyle}

\newcommand{\tr}{\mathop{{\rm Tr}}}

\newtheorem{definition}{Definition}[section]

\newtheorem{theorem}[definition]{Theorem}
\newtheorem{lemma}[definition]{Lemma}
\newtheorem{corollary}[definition]{Corollary}

\newtheorem{conjecture}[definition]{Conjecture}

\newcommand{\proof}{\noindent {\bf Proof:}\ }
\newcommand{\squ}{\noindent $\square$}

\numberwithin{equation}{section}

\begin{document}

\begin{titlepage}

\ \vskip 2.5cm

\centerline{\huge Domain Structure of Black Hole Space-Times}

\vskip 1.6cm
\centerline{\bf Troels Harmark}
\vskip 0.5cm
\centerline{\sl The Niels Bohr Institute}
\centerline{\sl Blegdamsvej 17, 2100 Copenhagen \O, Denmark}

\vskip 0.5cm

\centerline{\small\tt harmark@nbi.dk}

\vskip 1.6cm

\centerline{\bf Abstract} \vskip 0.2cm \noindent We introduce the
domain structure for stationary black hole space-times. Given a set
of commuting Killing vector fields of the space-time the domain
structure lives on the submanifold where at least one of the Killing
vector fields have zero norm. Depending on which Killing vector
field has zero norm the submanifold is naturally divided into
domains. A domain corresponds either to a set of fixed points of a
spatial symmetry or to a Killing horizon, depending on whether the
characterizing Killing vector field is space-like or time-like near
the domain. The domain structure provides invariants of the
space-time, both topological and geometrical. It is defined for any
space-time dimension and any number of commuting Killing vector
fields. We examine the domain structure for asymptotically flat
space-times and find a canonical form for the metric of such
space-times. The domain structure generalizes the rod structure
introduced for space-times with $D-2$ commuting Killing vector
fields. We analyze in detail the domain structure for Minkowski
space, the Schwarzschild-Tangherlini black hole and the Myers-Perry
black hole in six and seven dimensions. Finally we consider the
possible domain structures for asymptotically flat black holes in
six and seven dimensions.


\end{titlepage}

\small
\tableofcontents
\normalsize
\setcounter{page}{1}

\section{Introduction and summary}

It has been realized in recent years that the dynamics of black
holes in dimension $D \geq 5$ is much richer than in four
dimensions. In four dimensions the famous uniqueness theorems
\cite{Israel:1967wq} state that given the asymptotic charges, $i.e.$
the mass, angular momentum and the electric and magnetic charges,
there is at most one available black hole solution, namely the
Kerr-Newman solution. In dimensions $D\geq 5$ there are instead a
number of available solutions given the asymptotic charges, as first
realized with the discovery of the black ring in five dimensions
\cite{Emparan:2001wn}. This naturally brings up the question of
whether one can find a general set of invariants, in addition to the
asymptotic charges, that characterize a black hole space-time for $D
\geq 5$.

In this paper we propose a set of invariants given a stationary
black hole space-time with any number of space-time dimensions and
any number of commuting Killing vector fields. We call this set of
invariants the {\sl domain structure}. The domain structure lives on
the submanifold where at least one of the Killing vector fields have
zero norm. Depending on which Killing vector field has zero norm the
submanifold is naturally divided into domains. A domain corresponds
either to a set of fixed points of a spatial symmetry or to a
Killing horizon, depending on whether the characterizing Killing
vector field is space-like or time-like near the domain.

The domain structure generalizes the so-called {\sl rod structure}
proposed in \cite{Harmark:2004rm} as the set of invariants
characterizing asymptotically flat black holes in five dimensions
which are solutions of vacuum Einstein equations and possess two
rotational Killing vector fields. For this class of solutions the
submanifold for which at least one of the Killing vector fields have
zero norm can in certain canonical coordinates be seen as a line.
This line is then divided into intervals called rods according to
which Killing vector field has a zero norm. Such rods of
five-dimensional black holes was first considered for generalized
Weyl solutions in \cite{Emparan:2001wk}. The proposal of
\cite{Harmark:2004rm} that the rod structure provides a
characterization of asymptotically flat black holes in five
dimensions which are solutions of vacuum Einstein equations and
possess two rotational Killing vector fields is supported by the
uniqueness theorem of \cite{Hollands:2007aj} which states that a
black hole space-time with a single event horizon is unique given
the rod structure and the asymptotic charges.

The domain structure provides in particular a generalization of the
rod structure for the case of five-dimensional black hole
space-times with two rotational Killing vector fields (and more
generally solutions with $D-2$ commuting linearly independent
Killing vector fields) since in the approach of this paper we can
analyze solutions with matter fields such as gauge fields and scalar
fields. This overlaps with previous generalizations of the rod
structure \cite{Hollands:2007qf}. We reproduce furthermore the
constraints on the rod structure derived in \cite{Hollands:2007aj}.

The domain structure provides invariants of the black hole
space-time, both topological and geometrical. It reveals certain
aspects of the global structure of the black hole space-time. In
particular one can read off the topology of the event horizon(s). It
can also help in exploring what black hole space-times are possible.
The more commuting Killing vector fields one has, the more
invariants one obtains and the more constraints one finds on the
possible black hole space-times. In terms of topology of the event
horizon the topological censorship theorem says that it should be of
positive Yamabe type (i.e. it must admit a metric of positive
curvature) \cite{Galloway:2005mf}. If we have more than just the
Killing vector field associated with stationarity of the space-time
we can give further restrictions on the topology. Thus, not only we
can provide invariants to characterize the space-time, we can also
use the domain structure to provide limitations on what types of
black holes are possible. In addition to the topological invariants
the geometrical invariants which measures the volumes of each domain
can be used as a further characterization of the space-time.

We will always assume that our black hole space-time is stationary,
$i.e.$ it has a Killing vector field which is time-like far away
from the event horizon(s). If this is the only Killing vector field
the domain structure invariants will coincide with the previously
known topological data and physical parameters of the black hole
space-time. For example, the domain structure will give the topology
and the area of the event horizon(s). Given any number of additional
(asymptotically spatial) commuting Killing vector fields one finds
new invariants of the black hole space-time. For asymptotically flat
space-times the existence of at least one rotational Killing vector
field is guaranteed by the rigidity theorems of
\cite{Hollands:2006rj}.

We assume in this paper that the black hole space-times are either
asymptotically flat or asymptotically Kaluza-Klein space (here
defined as a $(D-q)$-dimensional Minkowski space times a
$q$-dimensional torus). Under this assumption we can find a
canonical form of the metric which is used to define the domain
structure. However, the general analysis of this paper can also work
for space-times with other asymptotics, such as asymptotically
Anti-de Sitter space-times. We save this generalization for a future
publication \cite{Harmark:domain2}.

More concretely the paper is built up as follows. In Section
\ref{sec:metric} we find a canonical form of the metric for all
asymptotically flat black hole space-times and all space-times which
are asymptotically Kaluza-Klein space.

In Section \ref{sec:domstruc} we define the domain structure for
black hole space-times. We analyze the structure of the kernel of
the metric on the commuting Killing vector fields and how this gives
rise to a hierarchy of submanifolds. We find that the submanifold
corresponding to the zero norm Killing vector fields is naturally
divided into domains and use this to define the domain structure. We
end with considering the special case of $D-2$ commuting Killing
vector fields where we obtain the rod structure now for a more
general class of solutions.

In Sections \ref{sec:sixdim} and \ref{sec:sevendim} we analyze the
Minkowski space, the Schwarzschild-Tangherlini black hole and the
Myers-Perry black hole \cite{Myers:1986un} in six and seven dimensions. These
space-times all possess $D-3$ commuting Killing vector fields. We
find coordinates such that the metric is put in a canonical form.
Using these coordinates we find the domain structure of the
space-times.

Finally we consider in Section \ref{sec:possible} which domain
structures are possible for asymptotically flat black hole
space-times in six and seven dimensions with $D-3$ commuting Killing
vector fields. Here we analyze in particular the domain structures
for the new types of black holes found recently using the Blackfold
approach \cite{Emparan:2007wm,Emparan:2009cs}. We also consider the
static numerical solutions recently found in \cite{Kleihaus:2009wh}.

We end in Section \ref{sec:concl} with discussing the implications
of the results of this paper and what new directions we can take. In
particular we discuss whether the domain structure along with the
asymptotic charges give enough invariants to fully characterize an
asymptotically flat black hole space-time. In line with this we
conjecture a uniqueness theorem for a certain class of black hole
space-times.

\section{Canonical form of metric}
\label{sec:metric}

We consider in the following a given $D$-dimensional space-time with
$p$ commuting linearly independent Killing vector fields. In detail
we are given a $D$-dimensional manifold $\CM_D$ with a Lorentzian
signature metric with $p$ commuting linearly independent Killing
vector fields $V_{(i)}$, $i=0,1,...,p-1$. The Killing vector fields
are such that they generate the isometry group $\R \times
U(1)^{p-1}$. In particular the $p-1$ $U(1)$ symmetries are generated
by the $p-1$ space-like Killing vector fields $V_{(i)}$,
$i=1,...,p-1$, while the Killing vector field $V_{(0)}$ generates
the $\R$ isometry. In the following we present the canonical form of
the metric for such a space-time.

\subsection{Preliminaries}
\label{sec:start}

As stated above we are given a $D$-dimensional space-time $\CM_D$
with $p$ commuting linearly independent Killing vector fields
$V_{(i)}$, $i=0,1,...,p-1$. Define $n= D-p$. We can always find a
coordinate system $x^0,x^1,...,x^{p-1},y^1,...,y^n$ such that
\begin{equation}
\label{killv} V_{(i)} = \frac{\partial}{\partial x^i}
\end{equation}
for $i=0,1,...,p-1$. We then write the metric as
\begin{equation}
\label{genmet} ds^2 = G_{ij} (dx^i + A^i_a dy^a) (dx^j + A^j_b dy^b)
+ \tilde{g}_{ab} dy^a dy^b
\end{equation}
with $i,j=0,1,...,p-1$ and $a,b=1,...,n$, and where $G_{ij}$,
$A^i_a$ and $\tilde{g}_{ab}$ only depend on $y^a$. In Appendix
\ref{sec:einstein} we compute the Ricci tensor for the metric
\eqref{genmet}.

The goal of this paper is to understand the structure of the kernel
of the metric $G_{ij}$ on the commuting Killing vector fields for
black hole space-times. If we have a point $q$ for which the kernel
$\ker G(q)$ is non-trivial it means that there exists (at least one)
Killing vector field $W$, which is a (non-zero) linear combination
of $V_{(i)}$, $i=0,1,...,p-1$, such that $G(q) W = 0$. Equivalently,
one can say that the norm of $W$, as measured with the metric
$G_{ij}$, is zero at $q$. Thus, the submanifold where at least one
linear combination of the Killing vector fields $V_{(i)}$ has zero
norm corresponds to the points where the kernel $\ker G$ is
non-trivial.

It is clear from this that whenever we have a point $q$ for which
$\ker G(q)$ is non-trivial we have that $\det G (q) = 0$. We
therefore use the function $\det G$ on the space-time to define one
of the coordinates in our canonical form for the metric.

Define now the $n$-dimensional manifold $\CN_n$ as the quotient
space $\CM_D / \sim$ where the equivalence relation $\sim$ is such
that two points in $\CM_D$ are equivalent if they can be connected
by an integral curve of a linear combination of the Killing vector
fields $V_{(i)}$, $i=0,1,...,p-1$. This is known as the {\sl orbit
space} of the space-time since each point in $\CN_n$ corresponds to
an orbit of the $\R \times U(1)^{p-1}$ symmetry of the commuting
Killing vector fields. The $n$-dimensional manifold $\CN_n$ is
naturally equipped with the $n$-dimensional part of the metric
\eqref{genmet}
\begin{equation}
\label{metnn} ds_n^2 =  \tilde{g}_{ab} dy^a dy^b
\end{equation}
where we sum over $a,b=1,2,...,n$.

On the $n$-dimensional manifold $\CN_n$
the fields $A^i_a$, $i=0,1,...,p-1$, can be thought of as components
of $p$ $U(1)$ gauge fields. Consider the coordinate transformation
$ x^i \rightarrow x^i - \alpha^i (y^a) $
with $y^a$ kept fixed. Under this coordinate transformation the
Killing vectors are still of the form \eqref{killv} and for the
metric \eqref{genmet} $G_{ij}$ and $\tilde{g}_{ab}$ stays the same
while $A^i_a$ transforms as $ A^i_a \rightarrow A^i_a + \partial
\alpha^i/\partial y^a$.
We see that this is a gauge transformation of the $U(1)$ gauge field
$A^i=A^i_a dy^a$.

In the following we would like to define new coordinates on the
$n$-dimensional manifold $\CN_n$ with metric \eqref{metnn} suitable
for examining the kernel of the metric $G_{ij}$ on the commuting
Killing vector fields $V_{(i)}$. To this end, define the function
$r(y^a)$ on $\CN_n$ as
\begin{equation}
\label{defrm} r^m \equiv \sqrt{ |\det G_{ij}| }
\end{equation}
for a positive real number $m$. We see then that the kernel of
$G_{ij}$ corresponds to $r=0$. We assume that $(\partial r /
\partial y^1 , ... ,\partial r / \partial y^n ) \neq 0$ on $\CN_n$
up to a subspace of $n$-volume zero. Define now the vector field
$\chi = \chi^a \partial / \partial y^a$ on $\CN_n$ by $\chi^a =
\tilde{g}^{ab} (\partial r/\partial y^b)$. Define the equivalence
relation $\sim$ on $\CN_n$ such that two points  are equivalent if
they are connected by an integral curve of $\chi$. We can then
consider the quotient space $\CN_n/ \sim$, which is an
$(n-1)$-dimensional manifold. Consider coordinates $z^1,...,z^{n-1}$
on $\CN_n/ \sim$. We can extend these coordinates to functions on
$\CN_n$. We see then that $z^\alpha$ is constant on the integral
curves of $\chi$ ($\alpha=1,2,...,n-1$). Therefore, $\chi^a
(\partial z^\alpha / \partial y^a)=0$. Clearly $r,z^1,...,z^{n-1}$
is a coordinate system on $\CN_n$ and $g^{rz^\alpha}=0$,
$\alpha=1,2,...,n-1$. We can thus write the metric on $\CN_n$ as
\begin{equation}
\label{partmet} ds_n^2 = e^{2A} dr^2 + \sum_{\alpha,\beta=1}^{n-1}
\tilde{\Lambda}_{\alpha \beta} dz^\alpha dz^\beta
\end{equation}
where $A(r,z^\alpha)$ and
$\tilde{\Lambda}_{\alpha\beta}(r,z^\gamma)$ are functions.

\subsection{Canonical form for particular class of metrics}
\label{sec:partclass}

Before treating the general class of metrics we first consider a
special class of metrics. This class consists of metrics solving the
vacuum Einstein equations $R_{\mu\nu} =0$. For a metric of the form \eqref{genmet} these equations can be written as
\eqref{ein1}-\eqref{ein3} in Appendix
\ref{sec:einstein}. We demand in addition that the $p$ Killing
vector fields are such that
\begin{equation}
\label{ortcondition} V_{(0)}^{[\mu_1} V_{(1)}^{\mu_2} \cdots
V_{(p-1)}^{\mu_{p}} D^\nu V_{(i)}^{\rho]}=0 \ \  \mbox{for all}\ \
i=0,1,...,p-1
\end{equation}

Among the space-times in this class is $D$-dimensional Minkowski
space, which in addition to the time-translation Killing vector
field has $[(D-1)/2]$ rotational Killing vector fields so
that we can take $p \leq 1+ [(D-1)/2]$.%
\footnote{Of course $D$-dimensional Minkowski space has $D$
commuting Killing vector fields but since our purpose here is to
study black hole space-times we are only interested in the
rotational and the time-translation Killing vector fields since
these are the only ones that can be shared by asymptotically flat
black hole solutions.} Also the Kaluza-Klein space $\R^{1,D-1-q}
\times T^q$ is in this class with $p \leq 1 + q + [(D-q-1)/2]$.

By Theorem \ref{ortform} of Appendix \ref{sec:einstein} the
condition \eqref{ortcondition} means we can write the metric as
\begin{equation}
\label{theortmetric}
ds^2 = G_{ij} dx^i dx^j + \tilde{g}_{ab} dy^a dy^b
\end{equation}
where we sum over $i,j=0,1,...,p-1$ and $a,b=1,2,...,n$ ($i.e.$ we
have $D=p+n$) and where $V_{(i)} = \partial / \partial x^i$,
$i=0,1,...,p-1$. We see that this corresponds to the metric
\eqref{genmet} with $A^i_a=0$. If we take the trace of
Eq.~\eqref{ein1} (with $A^i_a=0$) we get $\partial_a (
\sqrt{\tilde{g}} \tilde{g}^{ab} \partial_b r^m ) =0$ where $r(y^a)$
is defined in \eqref{defrm}. In the $(r,z^\alpha)$ coordinate system
introduced in Section \ref{sec:start} this gives $- \partial_r A +
\partial_r \log \sqrt{\det \tilde{\Lambda}_{\alpha\beta}} + (m-1)/r
=0$. We see that it is natural to choose $m=1$. Furthermore, we
define $\nu \equiv (n-1) A$ and $\Lambda_{\alpha\beta} \equiv \exp (
\frac{2A}{n-1} ) \tilde{\Lambda}_{\alpha\beta}$. With this the trace
equation $\partial_a ( \sqrt{\tilde{g}} \tilde{g}^{ab} \partial_b
r^m ) =0$ becomes
\begin{equation}
\label{rderlamb}
\partial_r \lambda = 0 \spa \lambda \equiv \sqrt{ | \det \Lambda_{\alpha\beta} | }
\end{equation}
One can show that $D$-dimensional Minkowski space $\R^{1,D-1}$ and
also the Kaluza-Klein space $\R^{1,D-1-q} \times T^q$ admit
$z^\alpha$ coordinates such that $\lambda=1$. We show this
explicitly for six and seven dimensional Minkowski space in Sections
\ref{sec:sixdim} and \ref{sec:sevendim}. Since for this particular
class of metrics we are only interested in asymptotically flat
solutions, or solutions that asymptote to Kaluza-Klein space, we can demand without loss of generality
that $\lambda \rightarrow 1$ for $r\rightarrow \infty$ with
$z^\alpha$ fixed. Hence from \eqref{rderlamb} it follows that
$\lambda=1$ everywhere.

In conclusion, we get that for any $D$-dimensional space-time
solving the vacuum Einstein equations with $p$ commuting linearly
independent Killing vector fields obeying \eqref{ortcondition} the
metric can be put in the form
\begin{equation}
\label{rzmetric}
ds^2 = G_{ij} dx^i dx^j + e^{2(n-1)\nu} dr^2 + e^{2\nu} \Lambda_{\alpha\beta} dz^\alpha dz^\beta \spa
r^2 = | \det G_{ij} | \spa \lambda = 1
\end{equation}
with the Killing vector fields given by \eqref{killv}. We call
\eqref{rzmetric} the {\sl canonical form of the metric}
for this class of space-times.

\subsection{Canonical form for general class of metrics}
\label{sec:genclass}

We now treat the general case. Thus, we consider here
$D$-dimensional space-times with $p$ commuting linearly independent
Killing vector fields $V_{(i)}$, $i=0,1,...,p-1$. The Killing vector
fields are such that they generate the isometry group $\R \times
U(1)^{p-1}$.

\subsubsection*{Asymptotic flatness or Kaluza-Klein space asymptotics}

From Section \ref{sec:start} we have that we can write the metric as
in Eq.~\eqref{genmet}. The $\CN_n$ part of the metric can
furthermore be written as in Eq.~\eqref{partmet} with $r$ defined in
\eqref{defrm}. Choosing $m=1$ and defining $\nu \equiv (n-1) A$ and $\Lambda_{\alpha\beta}
\equiv \exp ( \frac{2A}{n-1} ) \tilde{\Lambda}_{\alpha\beta}$ we see
that the metric on $\CN_n$ takes the form $e^{2(n-1)\nu} dr^2 + e^{2\nu} \Lambda_{\alpha\beta} dz^\alpha
dz^\beta$ with $r^2 = |\det G_{ij}|$.
Assuming furthermore that the space-time which we are considering
asymptote to either $D$-dimensional Minkowski space or Kaluza-Klein
space $\R^{1,D-1-q} \times T^q$ we can demand that
$\lambda \rightarrow 1$ for $r\rightarrow \infty$. Thus, we can
write the metric in the form
\begin{equation}
\label{rzmetric2}
\begin{array}{c} \ds ds^2 = G_{ij} (dx^i + A^i) (dx^j + A^j) +
e^{2(n-1)\nu} dr^2 + e^{2\nu} \Lambda_{\alpha\beta} dz^\alpha
dz^\beta
\\[2mm] \ds r^2 = | \det G_{ij} | \spa \lambda  \rightarrow 1 \
\mbox{for}\ r \rightarrow \infty
\end{array}
\end{equation}
with the Killing vector fields given by \eqref{killv} and $\lambda
\equiv \sqrt{|\det \Lambda_{\alpha\beta}|}$. We call
\eqref{rzmetric2} the {\sl canonical form of the metric} for
solutions asymptoting to $D$-dimensional Minkowski space or
Kaluza-Klein space $\R^{1,D-1-q} \times T^q$. This is a
generalization of the form \eqref{rzmetric} for solutions of the
vacuum Einstein equations obeying the condition
\eqref{ortcondition}. Instead here we can consider solutions which
couples to any type of matter fields, such as gauge fields or scalar
fields, since we do not use Einstein equations in getting the metric
\eqref{rzmetric2}.

If we consider transforming from the coordinates $y^a$ in
\eqref{genmet} to two different coordinate systems $(r,z^\alpha)$
and $(\tilde{r},\tilde{z}^\alpha)$ both making the metric to be of
the canonical form \eqref{rzmetric2} we immediately see from $r^2 =
|\det G_{ij}|$ that $r(y^a) = \tilde{r} (y^a)$ and hence
$\nu(y^a)=\tilde{\nu}(y^a)$. Considering now the transformation from
$(r,z^\alpha)$ to $(\tilde{r},\tilde{z}^\alpha)$ we see that
$\tilde{r} (r,z^\alpha) = r$. In general we can write
$\tilde{z}^\alpha (r,z^\beta)$. However, since $g^{rz^\alpha}=0$ we
have that $ g^{\tilde{r} \tilde{z}^\alpha } = e^{-2(n-1)\nu} (
\partial \tilde{z}^\alpha /\partial r)$. Therefore
$\tilde{z}^\alpha$ cannot depend on $r$ so the most general
transformation is $\tilde{z}^\alpha (z^\beta)$. Imposing furthermore
that $\lambda \rightarrow 1$ for $r \rightarrow \infty$ we get that
{\sl the only left over coordinate transformations in the canonical
form \eqref{rzmetric2} are $(n-1)$-volume preserving diffeomorphisms
of the $z^\alpha$ coordinates} (apart from rigid rotations of the
$x^i$ coordinates and the gauge transformations of $A^i$).

\subsubsection*{Other types of asymptotics}

We can generalize the above to include metrics which are not
asymptotically Minkowski space or Kaluza-Klein space. In general we
imagine having a background space-time $\CM_D^{(0)}$ that a given
class of space-times asymptotes to. For this background space-time
we can now find a function $\lambda_0(r,z^\alpha)$ such that
$\lambda/\lambda_0 \rightarrow 1$ for $r\rightarrow \infty$. So for
any space-time which asymptotes to $\CM_D^{(0)}$ we can write the
metric on the form
\begin{equation}
\label{rzmetric3}
\begin{array}{c} \ds
ds^2 = G_{ij} (dx^i + A^i) (dx^j + A^j) +
e^{2(n-1)\nu} dr^2 + e^{2\nu} \Lambda_{\alpha\beta} dz^\alpha
dz^\beta
\\[2mm] \ds
r^2 = | \det G_{ij} | \spa \frac{\lambda}{\lambda_0}
\rightarrow 1 \ \mbox{for}\ r \rightarrow \infty
\end{array}
\end{equation}
In this way we can for example treat asymptotically Anti-de Sitter
space-times. This will be considered in detail elsewhere
\cite{Harmark:domain2}. For asymptotically de Sitter space-times the
analysis proceeds differently since the asymptotic region includes
the cosmological horizon for which $r=0$ \cite{Harmark:domain2}.

\section{Domain structure}
\label{sec:domstruc}

In this section we introduce the domain structure for black hole
space-times. We focus here on asymptotically flat space-times and
asymptotically Kaluza-Klein space-times. This means the metric can
be put in the canonical form \eqref{rzmetric2}. The analysis is
straightforwardly generalizable to other classes of space-times as
well.

We consider in the following a $D$-dimensional manifold $\CM_D$ with
a Lorentzian signature metric with $p$ commuting linearly
independent Killing vector fields $V_{(i)}$, $i=0,1,...,p-1$. The
Killing vector fields are such that they generate the isometry group
$\R \times U(1)^{p-1}$. In particular the $p-1$ $U(1)$ symmetries
are generated by the $p-1$ space-like Killing vector fields
$V_{(i)}$, $i=1,...,p-1$, while the Killing vector field $V_{(0)}$
generates the $\R$ isometry.

For purposes of our analysis we assume below that the following two
regularity conditions on the metric \eqref{rzmetric2} are obeyed:
The $A^i_a$ fields and the scalar fields $V^\mu_{(i)} V^\nu_{(j)}
R_{\mu\nu}$ do not go to infinity for $r\rightarrow 0$.

This Section is built up as follows. In Section \ref{sec:flows} we
consider the flow of the $p$ Killing vector fields. In Section
\ref{sec:kerG} we examine how the kernel of the Killing metric gives
rise to a natural hierarchy of submanifolds of $\CN_n$. In Section
\ref{sec:domains} we define the domains and their directions and use
this to define the domain structure. Finally in Section
\ref{sec:rodstruc} we discuss the reduction of the domain structure
to the rod structure in the special case of $n=2$.

The analysis of this section builds on generalizations of methods
used in Refs.~\cite{Harmark:2004rm} and \cite{Hollands:2007aj}.

\subsection{The flow of the Killing vector fields}
\label{sec:flows}

Before considering the form of the metric we consider here the flow
of the Killing vector fields. This will be of importance for the
analysis below. We define the flow of $V_{(i)}$ as
\begin{equation}
\sigma_s^{(i)} ( x^0 , ... , x^i , .... , x^{p-1} , y^1 ,..., y^n )
= ( x^0 , ... , x^i + s , .... , x^{p-1} , y^1 ,..., y^n )
\end{equation}
for $i=0,1,...,p-1$. For $i=1,2,...,p-1$ the Killing vector field
$V_{(i)}$ generates a $U(1)$ isometry hence the flow is periodic. We
normalize the periods of the flows with $i=1,2,...,p-1$ to be
$2\pi$. Note that for $i=1,2,...,p-1$ the Killing vector fields
$V_{(i)}$ can not be time-like anywhere since then one would have
closed time-like curves.

The set of Killing vector fields $V_{(i)}$, $i=0,1,...,p-1$,
corresponds to a particular choice of basis. We are not entirely
free to choose any basis. Consider the space-like Killing vectors
$V_{(i)}$, $i=1,...,p-1$. A new basis $W_{(i)}$, $i=1,...,p-1$ is in
general a linear combination
\begin{equation}
W_{(i)} = \sum_{j=1}^{p-1} U_{ij} V_{(j)}
\end{equation}
Considering in particular $W_{(1)}$ this generates the flow
\begin{equation}
\label{Wflow} \sigma_s ( x^0 , x^1 , x^2 , .... , x^{p-1} , y^1
,..., y^n ) = ( x^0 , x^1 + U_{11} s , x^2 + U_{12} s, .... ,
x^{p-1} + U_{1,p-1} s, y^1 ,..., y^n )
\end{equation}
We want $W_{(1)}$ to generate a $U(1)$ isometry and we choose the
period of the flow to be $2\pi$. From \eqref{Wflow} we see this
means that the $U_{1i}$ entries should be relatively prime numbers.
Thus we get the general requirement that $U \in GL(p-1, \Z)$ and
that the rows of $U$ should be relatively primes.

\subsection{Submanifolds and the kernel of $G$}
\label{sec:kerG}

In the following we would like to examine the structure of the
kernel $\ker G$ of the metric $G_{ij}$ on the commuting Killing
vector fields $V_{(i)}$. As stated above, if we have a point $q$ for
which the kernel $\ker G$ is non-trivial then $\det G = 0$ at $q$.
 We define therefore the set
\begin{equation}
\label{Bset}
B = \{ q \in \CN_n | \det G (q) = 0 \}
\end{equation}
This is a codimension one hypersurface in $\CN_n$, $i.e.$ it is an
$(n-1)$-dimensional submanifold. As we shall see in the following
this can be seen as part of the boundary of the manifold $\CN_n$.
Note that $B$ need not be a connected set (see examples in Section
\ref{sec:possible}). In the canonical form for the metric
\eqref{rzmetric2} $B$ is the set of points with $r=0$. In this
coordinate system we can naturally equip $B$ with the metric%
\footnote{Note that for space-times solving the vacuum Einstein
equations and obeying \eqref{ortcondition} this metric has
determinant $\lambda=1$.}
\begin{equation}
\label{Bmet}
ds_B^2 = \Lambda_{\alpha\beta}|_{r=0} dz^\alpha dz^\beta
\end{equation}

In the following we discuss for a given point $q\in B$ the Killing vector fields $v\in \ker G(q)$. We distinguish between Killing vector fields $v$ which are space-like (time-like) near $q$, meaning that there exists a neighborhood $\CO$ of $q$ with respect to the manifold $\CN_n$ such that $v^2 > 0$ ($v^2 < 0$) for any point in $\CO - B$.

Define now the sets
\begin{equation}
Q_k = \{ q \in \CN_n | \dim \ker G (q) \geq k \}
\end{equation}
for $k \in \{ 0,1,...,p \}$. Clearly $Q_0 = \CN_n$ and $Q_1 = B$. We examine now the properties of the sets $Q_k$.

\begin{theorem}
\label{theo:codim}
Consider a point $q \in Q_k - Q_{k+1}$. Then $Q_k$ is a codimension $k$ submanifold of $\CN_n$ in a neighborhood of $q$.

\proof Since $q \in Q_k - Q_{k+1}$ we have $k$ linearly independent Killing vectors fields $W_{(i)} \in \ker G$, $i=1,2,...,k$. We can always assume that at most one of these Killing vectors are time-like near $q$. If there were two of them which are time-like near $q$ it follows from the fact that $V_{(1)},...,V_{(p-1)}$ are space-like everywhere outside $B$ that we can form a linear combination of the two Killing vectors which is space-like near $q$.

Assume first that all of these Killing vector fields are space-like near $q$. Then in order to avoid a conical singularity each of these Killing vector fields should generate a $U(1)$ isometry (see Section \ref{sec:flows} for conditions on this, in particular one can infer from here that the $k$ Killing vector fields are everywhere space-like). We can now find Riemannian Normal Coordinates $n^0,n^1,...,n^{D-1}$ in a neighborhood of $q$ such that at $q$ the $D$-dimensional metric is $ds^2 = \eta_{\mu\nu} dn^\mu dn^\nu$, the first derivatives of the metric at $q$ in this coordinate system are zero, and such that the $k$ Killing vectors can be written as %
\footnote{A similar construction appeared in
\cite{Hollands:2007aj}.}
\begin{equation}
W_{(i)} = n^{2i} \frac{\partial}{\partial n^{2i+1}} -  n^{2i+1} \frac{\partial}{\partial n^{2i}} \ , \ i=1,2,...,k
\end{equation}
That is, $W_{(i)}$ is the rotational Killing vector field in the
plane $(n^{2i},n^{2i+1})$. Consider now the radii $\rho_i \equiv
\sqrt{(n^{2i})^2+(n^{2i+1})^2}$, $i=1,2,...,k$. First we observe
that putting any of the $\rho_i > 0$ we have $\dim \ker G < k$.
Secondly, for any point in the neighborhood of $q$ we see that if
$\rho_i=0$ for all $i=1,2,...,k$ then we are in $Q_k$. Therefore we
have shown that $Q_k = \{ q' \in \CN_n | \rho_i(q') =0 \ \forall
i=1,2,...,k \}$ in a neighborhood of $q$. Thus, we conclude that
$Q_k$ is a codimension $k$ submanifold of $\CN_n$ in a neighborhood
of $q$.

Assume now that $W_{(k)}$ is time-like near $q$ while $W_{(1)},...,W_{(k-1)}$ all are space-like near $q$. Then $W_{(k)}$ generates $\R$ while $W_{(i)}$ generate $U(1)$ for $i=1,2,...,k-1$. We can now find Riemannian Normal Coordinates $n^0,n^1,...,n^{D-1}$ in a neighborhood of $q$ such that at $q$ the $D$-dimensional metric is $ds^2 = \eta_{\mu\nu} dn^\mu dn^\nu$, the first derivatives of the metric at $q$ in this coordinate system are zero, and such that the $k$ Killing vectors can be written as
\begin{equation}
W_{(k)} = n^{0} \frac{\partial}{\partial n^{1}} +  n^{1} \frac{\partial}{\partial n^{0}} \spa W_{(i)} = n^{2i} \frac{\partial}{\partial n^{2i+1}} -  n^{2i+1} \frac{\partial}{\partial n^{2i}} \ , \ i=1,2,...,k-1
\end{equation}
As in the other case we have the radii $\rho_i \equiv \sqrt{(n^{2i})^2+(n^{2i+1})^2}$, $i=1,2,...,k-1$. We see that $W_{(k)}$ is the time-like Killing vector field in the Rindler space given by the coordinates $(n^0,n^1)$. We can therefore define the distance from the horizon in Rindler space as $\rho_k = \sqrt{(n^1)^2- (n^0)^2 }$. Proceeding with the argument the same way as above we see that $Q_k = \{ q' \in \CN_n | \rho_i(q') =0 \ \forall i=1,2,...,k \}$ in a neighborhood of $q$ and hence that $Q_k$ is a codimension $k$ submanifold of $\CN_n$ in a neighborhood of $q$.
\squ
\end{theorem}

From this theorem we can infer the following corollary:

\begin{corollary}
\label{cor:qsub}
Consider a point $q \in Q_{k+1} - Q_{k+2}$. Then $Q_{k+1}$ is a codimension one submanifold of $Q_k$ in a neighborhood of $q$.
\squ
\end{corollary}

We conclude from the above analysis that the structure of $\ker G$ naturally give rise to a hierarchy of submanifolds $Q_k$.

\subsection{Domains and their directions}
\label{sec:domains}

Suppose now that we consider a point $q \in B - Q_2$. Let
furthermore $D \subset B - Q_2$ be the maximally possible connected
set containing $q$. Write the coordinates for the point $q$ as
$(r,z^\alpha)=(0,z_*^\alpha)$. Since $\dim \ker G (q) = 1$ we can
find a Killing vector field $W \in \ker G(q)$.

Suppose $W$ is a space-like Killing vector field near $q$. In the
following we aim to show that the linear space $\ker G$ is constant
over $D$, $i.e.$ that $W \in \ker G (q')$ for any point $q' \in D$.

One way to show the constancy of $\ker G$ over $D$ is as follows \cite{Harmark:2004rm}. We
first observe that we can rigidly rotate $G_{ij}$ such that $W =
\partial /
\partial x^1$. For $r\rightarrow 0$ and $z^\alpha \rightarrow
z_*^\alpha$ we then have $G_{11} = c^2 r^2 + \CO (r^3)$ with $c$ a
constant, and the entries of $G_{ij}$ with $i,j\neq 1 $ approaching
a constant. In order for the space-time to be regular near $q$ we
need that $g_{rr} = e^{2(n-1)\nu}$ approaches a non-zero constant.
We write this as $\nu \rightarrow c'$ for $r\rightarrow 0$ and
$z^\alpha \rightarrow z_*^\alpha$. We also need that $A^1
\rightarrow 0$ for $r\rightarrow 0$ and $z^\alpha \rightarrow
z_*^\alpha$. The metric \eqref{rzmetric2} thus approaches
\begin{equation}
\label{neardom} ds^2 = c^2 r^2 (dx^1)^2 + e^{2(n-1)c'} dr^2 +
\sum_{i,j\neq 1} G_{ij}|_{q} \, (dx^i+A^i) (dx^j+A^j) + e^{2c'}
\Lambda_{\alpha\beta}|_{q} \, dz^\alpha dz^\beta
\end{equation}
for $r\rightarrow 0$ and $z^\alpha \rightarrow z_*^\alpha$. Consider
now the $R_{ij}$ part of the Ricci tensor \eqref{ricci_ij}. From
requiring regularity we have that $R_{ij} = V^\mu_{(i)} V^\nu_{(j)}
R_{\mu\nu}$ should be finite for $r\rightarrow 0$. Examining now
$R_{ii}$ for $i\neq 1$ one finds that $\tilde{g}^{ab}
\partial_a G_{1i} \partial_b G_{1i} \rightarrow 0$ for $r\rightarrow
0$ and $z^\alpha \rightarrow z_*^\alpha$. This gives that
$\partial_a G_{1i} = 0$ in $q$. Picking now any other point in $D$
we can do the same. Since $D$ is connected this means that $W =
\partial / \partial x^1$ is in $\ker G$ everywhere in $D$.
Undoing the rigid rotation we have shown that if $W \in \ker G(q)$
then $W \in \ker G(q')$ for any point $q' \in D$.

Another way to show the constancy of $\ker G$ over $D$ is to employ
the fact that to have a regular metric at $q$ we need that $W$
generates a $U(1)$ isometry \cite{Hollands:2007aj}. Otherwise we get
a conical singularity. This means we have
\begin{equation}
W = \sum_{i=1}^{p-1} q_i V_{(i)}
\end{equation}
where the $q_i$'s are rational numbers. Since the norm of $W$ is not
significant we can choose to restrict the $q_i$'s to be relatively
prime numbers. Now, for every point of $q' \in D$ we have an
eigenvector $W_{q'} \in \ker G(q')$. But since $W_{q'}$ should vary
continuously over $D$ we see that one necessarily must have that
$W_{q'} = W$. Thus we can conclude that $W \in \ker G$ everywhere in
$D$ hence $\ker G$ is constant over $D$.

Suppose instead $W$ is a time-like Killing vector field near $q$.
Making a rigid rotation of $G_{ij}$ we can put $W = \partial /
\partial x^0$. For the space-time to be regular near $p$ the metric
should approach
\begin{equation}
\label{neardom2} ds^2 = - c^2 r^2 (dx^0)^2 + e^{2(n-1)c'} dr^2 +
\sum_{i,j\neq 0} G_{ij}|_{q} \, (dx^i+A^i) (dx^j+A^j) + e^{2c'}
\Lambda_{\alpha\beta}|_{q} \, dz^\alpha dz^\beta
\end{equation}
for $r\rightarrow 0$ and $z^\alpha \rightarrow z_*^\alpha$ where $c$
and $c'$ are constants. Just as above one can examine $R_{ii}$ for
$i\neq 0$ using \eqref{ricci_ij} and find that $\partial_a G_{0i} =
0$ in $p$. Since $p$ is any point in $D$ this means that $W =
\partial / \partial x^0 \in \ker G$ everywhere in $D$.

It follows furthermore from the fact that $W = \partial / \partial x^0 \in \ker G$ everywhere in $D$ and the near-$q$ metric \eqref{neardom2} that for $r\rightarrow 0$ and $z^\alpha \rightarrow z^\alpha_*$ with $(0,z_*^\alpha) \in D$ the metric approaches a Rindler space-time times a regular space of Euclidean signature. Therefore $D$ is a Killing horizon of the Killing vector field $W = \partial / \partial x^0$.

We have thus shown above that the vectors in $\ker G$ are constant in the connected pieces of $B- Q_2$. With this, we can make the following definition:

\begin{definition}
Let $q \in B- Q_2$ and let $W \in \ker G(q)$. A domain $D$ containing $q$ is the maximal connected set in $B$ such that $q\in D$ and such that for any point $q' \in D$ we have $W \in \ker G(q')$. \squ
\end{definition}

We can now consider all the distinct domains of $B$, write them as $D_1,...,D_N$. Clearly we have $D_i \cap D_j \subset Q_2$. From Corollary \ref{cor:qsub} we have that $Q_2$ can be seen locally as a submanifold
of $B$ of codimension one. This means that for $q \in Q_2$ any neighborhood of $q$ in $B$ will contain points in $B- Q_2$. This shows that the domains of $B$ contains all points in $Q_2$. Thus we can write
$B = D_1 \cup D_2 \cup \cdots \cup D_N$.

We have now shown the following theorem:

\begin{theorem}
\label{theo:dom} Let $D_1,...,D_N$ be the domains of $B$. We have
that $B = D_1 \cup D_2 \cup \cdots \cup D_N$. For each domain $D_m$
we can find a Killing vector field $W_m$ such that $W_m \in \ker G$
for all points in $D_m$. We call $W_m$ the {\sl direction} of the
domain $D_m$. If $W_m$ is space-like for $r\rightarrow 0$ we can
write it in the form
\begin{equation}
\label{Wmspace}
W_m = \sum_{i=1}^{p-1} q_i V_{(i)}
\end{equation}
where the $q_i$'s are relatively prime numbers. Then $W_m$ generates
a $U(1)$ isometry and the generated flow has period $2\pi$. In this
case we say that the direction $W_m$ is space-like.

If $W_m$ is time-like for $r\rightarrow 0$ we can write it in the
form
\begin{equation}
\label{Wmtime}
W_m = V_{(0)} +  \sum_{i=1}^{p-1} \Omega_i V_{(i)}
\end{equation}
and the domain $D_m$ is a Killing horizon for the Killing vector field $W_m$. In this case we say that the direction $W_m$ is time-like.
\squ
\end{theorem}

From Theorem \ref{theo:dom} we can now define the domain structure of a solution:

\begin{definition}
\label{def:domstruc}
The domain structure of a solution is defined as the split-up of $B$ in domains $B = D_1 \cup D_2 \cup \cdots \cup D_N$ up to volume preserving diffeomorphisms, along with the directions $W_m$, $m=1,2,...,N$, of the domains.
\end{definition}

Our results above show that the domain structure of a given solution
gives invariants of the solution (up to rigid transformations of the
Killing vector fields as discussed in Section \ref{sec:flows}). In
particular we have shown in Section \ref{sec:genclass} that the only
left over coordinate transformations in the $(r,z^\alpha)$
coordinates are volume-preserving diffeomorphisms of the $z^\alpha$
coordinates.

Since the domain structure of a solution gives invariants of the
solution it can characterize the solution. That is, it gives
invariants that can help in distinguishing different solutions, and
it can furthermore provide information about the nature of the
difference. A particular set of invariants derived from the domain
structure consists of the volumes of the domains with respect to the
metric \eqref{Bmet}. We call these invariants geometrical since they
define in a coordinate-invariant way sizes of well-defined regions
of the space-time as measured with the metric of the space-time.

In Section \ref{sec:concl} we discuss for which type of solutions we
can expect the domain structure to give a full characterization. We
conjecture a uniqueness theorem for this type of solutions. We also
discuss what extra information one has to add beyond the domain
structure to give a full characterization of solutions coupled to
gauge fields.

\subsection{Reduction to the rod structure for $n=2$}
\label{sec:rodstruc}

We consider here the special case $n=2$, $i.e.$ with $p=D-2$ Killing
vector fields. This is the case studied in
\cite{Harmark:2004rm,Hollands:2007aj,Hollands:2007qf}.

\subsubsection*{Solutions of vacuum Einstein equations}

We consider first solutions of vacuum Einstein equations with
$p=D-2$ commuting Killing vector fields. By theorem \ref{usualcase}
this means (under mild assumptions) that we can put the metric in
the canonical form \eqref{rzmetric} which in this case reduces to
\begin{equation}
ds^2 = G_{ij} dx^i dx^j + e^{2\nu} ( dr^2 + dz^2 ) \spa r^2 = |\det
G_{ij}|
\end{equation}
which is the canonical form of the metric found in
\cite{Harmark:2004rm}. Assuming $\CN_2$ is simply connected we have
$B = \R$, $i.e.$ it is the $z$-axis for $r=0$. Let $D_1,...,D_N$ be
the domains of $B$ with directions $W_1,...,W_N$. Then each domain
corresponds to an interval of the $z$-axis. Thus, in the
nomenclature of \cite{Harmark:2004rm} each domain corresponds to a
rod. Furthermore the direction of the rod is simply the direction of
the domain. We also see that the volume preserving diffeomorphisms
mentioned in Definition \ref{def:domstruc} here simply are the
translations. The fact that the volumes of the domains are
invariants corresponds to the statement that the lengths of the rods
are invariants.

We thus regain the rod structure of \cite{Harmark:2004rm}. We found
moreover that the directions of the space-like rods can be written
as \eqref{Wmspace} with the $q_i$ being relatively prime numbers.
This constraint has previously been found in \cite{Hollands:2007aj}.

\subsubsection*{General case}

For the more general case of asymptotically flat or asymptotically
Kaluza-Klein space solutions with $n=2$ we get from
\eqref{rzmetric2} that the canonical form of the
metric is
\begin{equation}
\begin{array}{c} \ds
ds^2 = G_{ij} (dx^i + A^i) (dx^j + A^j) + e^{2\nu} ( dr^2 +
\lambda^2 dz^2 )\spa r^2 = |\det G_{ij}|
\\[2mm] \ds
\lambda \rightarrow 1 \ \mbox{for}\ r \rightarrow \infty
\end{array}
\end{equation}
This is more general than the form found in
\cite{Harmark:2004rm,Hollands:2007aj,Hollands:2007qf} since here we
are not specific on what kinds of matter fields appear in the
solution. Other than that the domains/rods are again intervals on
the $z$-axis defined by $r=0$. The lengths of the domains/rods are
measured by the metric
\begin{equation}
ds^2 = \lambda^2 |_{r=0} dz^2
\end{equation}
These lengths are invariants of the black hole space-time. The
domain/rod structure is defined up to translations. We have thus
defined the rod structure for any asymptotically flat or
asymptotically Kaluza-Klein black hole space-time with $p=D-2$
commuting Killing vector fields. We can furthermore extend the rod
structure to include non-asymptotically flat solutions
\cite{Harmark:domain2}.

\section{Domain structure of six dimensional black holes}
\label{sec:sixdim}

In this section we analyze the known asymptotically flat
six-dimensional exact solutions of the vacuum Einstein equations.
These are the Minkowski space, the Schwarzschild-Tangherlini black
hole and the Myers-Perry black hole. They all have three Killing
vector fields, which is the maximally possible number in six
dimensions. In addition the Killing vector fields obey the condition
\eqref{ortcondition}. This means that the metrics can be put in the
canonical form \eqref{rzmetric} with $p=n=3$.

\subsubsection*{Minkowski space}

The metric of six-dimensional Minkowski space is
\begin{equation}
ds^2 = - dt^2 + d\rho^2 + \rho^2 ( \mu_1^2 d\phi_1^2 + \mu_2^2
d\phi_2^2
 +  d\theta^2 +  \cos^2 \theta d\psi^2 )
\end{equation}
with
\begin{equation}
\label{dircos6D} \mu_1 = \sin \theta \spa \mu_2 = \cos \theta \sin
\psi \spa \mu_3 = \cos \theta \cos \psi
\end{equation}
and with the coordinate ranges $0 \leq \theta \leq \pi/2$ and $0
\leq \psi \leq \pi$. From \eqref{rzmetric} we see
\begin{equation}
r = \rho^2 \mu_1 \mu_2 = \rho^2 \sin \theta \cos \theta \sin \psi =
\frac{1}{2} \rho^2 \sin  ( 2 \theta ) \sin \psi
\end{equation}
Using this we get $e^{-4\nu} = \rho^2 \big[ \sin^2\psi + \sin^2 \theta
\cos^2 \psi \big]$.
In order to fit in the canonical form \eqref{rzmetric}
we need to find the $z^\alpha$ coordinates such that
$g_{rz^\alpha}=0$ and $\lambda=1$. We make the following ansatz
$z^\alpha = \rho^{k_\alpha} F_\alpha (\theta) ( \cos \psi
)^{l_\alpha}$,
$\alpha=1,2$. Demanding that $g_{rz^\alpha}=0$ gives that the
functions $F_\alpha(\theta)$ are of the form $F_\alpha(\theta) = C_\alpha (\cos \theta)^{l_\alpha} ( \cos 2\theta
)^{\frac{k_\alpha-l_\alpha}{2}}$
where $C_\alpha$ are constants. One can furthermore infer that
$\lambda=1$ provided $C_1 C_2 = \pm 1/(k_1 l_2 - k_2 l_1)$, $k_1+k_2 = 3 $ and $l_1+l_2 = 1$.
We choose therefore the coordinates
\begin{equation}
\label{flatrzz} r = \frac{1}{2} \rho^2 \sin 2\theta \sin \psi \spa
z^1 = \rho \cos \theta \cos \psi \spa z^2 = \frac{1}{2} \rho^2 \cos
2 \theta
\end{equation}
With this choice of coordinates the 6D flat space metric is put in
the form \eqref{rzmetric}.

We now analyze the domain structure of six-dimensional Minkowski
space using the coordinates \eqref{flatrzz}. This can be done by
analyzing the coordinates $z^\alpha$ when $r=0$. We find the domain
structure
\begin{equation}
\label{6Dminkdomains}
\begin{array}{c} \ds
W_1 = \frac{\partial}{\partial \phi_1} \spa D_1 = \big\{ (z^1,z^2)
\in \R^2 \big| z^2 \geq \frac{1}{2} (z^1)^2 \big\} \\[4mm] \ds
W_2 =
\frac{\partial}{\partial \phi_2} \spa D_2 = \big\{ (z^1,z^2) \in
\R^2 \big| z^2 \leq \frac{1}{2} (z^1)^2 \big\}
\end{array}
\end{equation}
We see that $D_1 \cup D_2 = \R^2$. This domain structure is depicted
in the top left diagram of Figure \ref{domplots}.
We note that in terms of the $(\rho,\theta,\psi)$ coordinates the
two domains correspond to $D_1: \theta=0$ and $D_2: \psi=0,\pi$.

Building on our parametrization of six-dimensional Minkowski space
\eqref{flatrzz} we can now describe the boundary conditions that we
wish to impose on six-dimensional asymptotically flat space-times.
We consider here solutions with $p=3$ such that one can write them
in the form \eqref{rzmetric2} in terms of coordinates $(r,z^1,z^2)$.
We define the asymptotic region in $(r,z^1,z^2)$ coordinates as $L
\rightarrow \infty$ with $\sqrt{r}/L$, $(z^1)^2/L$ and $z^2/L$
finite or going to zero where $L \equiv  r + (z^1)^2 + |z^2|$. In
this asymptotic region we require that the metric should asymptote
to six-dimensional Minkowski space. This means in particular that we
require the $(r,z^1,z^2)$ coordinates to asymptote to
Eq.~\eqref{flatrzz}. For the domain structure at $r=0$ this means
that for $(z^1)^2 + |z^2| \rightarrow \infty$ we have the two
domains \eqref{6Dminkdomains}, with the border between the domains
being at the curve $z^2 = (z^1)^2 /2 $ up to corrections of order
$((z^1)^2 + |z^2|)^{-1/2}$.

\subsubsection*{Schwarzschild-Tangherlini black hole}

The 6D Schwarzschild-Tangherlini black hole has the metric
\begin{equation}
ds^2 = - f dt^2 + \frac{d\rho^2}{f} + \rho^2 ( \mu_1^2 d\phi_1^2 +
 \mu_2^2 d\phi_2^2  +  d\theta^2 + \cos^2 \theta
d\psi^2 ) \spa f = 1 - \frac{\rho_0^3}{\rho^3}
\end{equation}
with the director cosines given by \eqref{dircos6D}. We have from
\eqref{rzmetric}
\begin{equation}
\label{schw6r} r = \rho^2 \sqrt{f} \mu_1 \mu_2 = \rho^2 \sqrt{f}
\sin \theta \cos \theta \sin \psi = \frac{1}{2} \rho^2 \sqrt{f} \sin
( 2 \theta ) \sin \psi
\end{equation}
From this one can easily compute $\exp(-4\nu)$ as function of
$(\rho,\theta,\psi)$. We need to impose that $g_{rz^\alpha}=0$ and $\lambda=1$. Make now
the ansatz
$z^1 = b_1(\rho) \cos \theta \cos \psi$ and $z^2 = b_2(\rho) \cos
2\theta$.
Then $g_{rz^\alpha}=0$ is equivalent to $2 b_1'/b_1 = b_2' / b_2 = 8\rho^3 /(4\rho^4-\rho \rho_0^3)$.
We therefore get the $z^\alpha$ coordinates
\begin{equation}
\label{schw6z} z^1 = \rho \Big( 1 - \frac{\rho_0^3}{4\rho^3}
\Big)^{\frac{1}{3}} \cos \theta \cos \psi  \spa z^2 = \frac{1}{2}
\rho^2 \Big( 1 - \frac{\rho_0^3}{4\rho^3} \Big)^{\frac{2}{3}} \cos
2\theta
\end{equation}
Comparing this with \eqref{flatrzz} for six-dimensional Minkowski
space we see that we have the right asymptotic behavior, as
discussed above. One can also compute that $\lambda=1$ which indeed
is guaranteed by Eq.~\eqref{rderlamb} and by choosing the right
asymptotics.

The domain structure for the six-dimensional
Schwarzschild-Tangherlini black hole as found from the coordinates
\eqref{schw6r} and \eqref{schw6z} is given by
\begin{equation}
\begin{array}{c} \ds
W_1 = \frac{\partial}{\partial \phi_1} \spa D_1 = \Big\{ (z^1,z^2)
\in \R^2 \Big| z^2 \geq K \spa z^2 \geq \frac{1}{2} (z^1)^2
\Big\} \\[4mm] \ds
W_2 = \frac{\partial}{\partial \phi_2} \spa D_2 = \Big\{ (z^1,z^2)
\in \R^2 \Big| z^2 \leq (z^1)^2 - K \spa z^2 \leq \frac{1}{2}
(z^1)^2 \Big\}
\\[4mm] \ds
W_3 = \frac{\partial}{\partial t} \spa D_3 = \Big\{ (z^1,z^2) \in
\R^2 \Big| (z^1)^2 - K \leq z^2 \leq K \Big\}
\end{array}
\end{equation}
where we defined the constant $K \equiv (\rho_0^2/2) (3/4)^{2/3}$.
This domain structure is depicted
in the middle left diagram of Figure \ref{domplots}. We note that in terms of the $(\rho,\theta,\psi)$ coordinates the
three domains correspond to $D_1: \theta=0$, $D_2: \psi=0,\pi$ and
$D_3: \rho=\rho_0$.

\subsubsection*{Myers-Perry black hole}

The six-dimensional Myers-Perry black hole solution is \cite{Myers:1986un}
\begin{equation}
ds^2 = - dt^2 + \sum_{i=1}^2 (\rho^2 + a_i^2) ( d\mu_i^2 + \mu_i^2
d\phi_i^2 ) + \rho^2 d\mu_{3}^2 + \frac{\rho_0^{3} \rho }{\Pi F}
\Big( dt - \sum_{i=1}^2 a_i \mu_i^2 d\phi_i \Big)^2 + \frac{\Pi F
d\rho^2}{\Pi - \rho \rho_0^{3} }
\end{equation}
Here the director cosines are given by \eqref{dircos6D} and we have
\begin{equation}
F ( \rho , \mu_i ) = 1 - \sum_{i=1}^2 \frac{a_i^2 \mu_i^2}{\rho^2 +
a_i^2 } \spa \Pi(\rho) = \prod_{i=1}^2 ( \rho^2 + a_i^2 )
\end{equation}
The horizon is placed at $\rho=\rho_h$ which is defined as the
largest real root of the equation $\Pi (\rho) = \rho \rho_0^3$. We
find from \eqref{rzmetric}
\begin{equation}
\label{therfullMP} r = \sqrt{\Pi - \rho \rho_0^3}\, \mu_1 \mu_2 =
\frac{1}{2} \sqrt{\Pi - \rho \rho_0^3}\, \sin ( 2\theta ) \sin \psi
\end{equation}
We make the ansatz
\begin{equation}
\label{zz6DMP} z^1 = b_1(\rho) \cos \theta \cos \psi \spa z^2 =
b_2(\rho) \cos 2\theta + p(\rho) \cos^2 \theta \cos^2 \psi
\end{equation}
Demanding $g_{rz^\alpha}=0$ is equivalent to the equations
\begin{equation}
\label{peq} \begin{array}{c} \ds
\frac{b_1'}{b_1} = \frac{4\rho^2 + 2(a_1^2+a_2^2)}{4\rho^3 + 2\rho
(a_1^2+a_2^2) - \rho_0^3} \spa \frac{b_2'}{b_2} =
\frac{8\rho^2}{4\rho^3 +2\rho(a_1^2 +a_2^2 ) - \rho_0^3 }
\\[5mm] \ds
(4\rho^3 +2\rho(a_1^2 +a_2^2 ) - \rho_0^3) p' -
4(2\rho^2 + a_1^2 + a_2^2 ) p = 4 a_2^2 b_2
\end{array}
\end{equation}
Imposing the boundary conditions for $\rho \rightarrow \infty$ we
get
\begin{equation}
\label{bsols} \begin{array}{c} \ds b_1 (\rho) = \rho  \exp \left\{ -
\int_{\rho/\rho_0}^\infty \frac{dx}{x \left(4x^3+2 x A^2 - 1
\right)} \right\} \\[6mm] \ds  b_2 (\rho) = \frac{1}{2} \rho^2 \exp \left\{ -
\int_{\rho/\rho_0}^\infty \frac{\left( 2 - 4 x A^2 \right)dx}{x
\left(4x^3+ 2 x A^2 - 1 \right)} \right\}
\end{array}
\end{equation}
with $A^2 \equiv (a_1^2+a_2^2)/\rho_0^2$.
Considering the last equation in \eqref{peq} we see that this is
solved by
\begin{equation}
\label{psol} p = \frac{a_2^2 }{a_1^2 + a_2^2 } ( b_1^2 - 2 b_2 )
\end{equation}
where we fixed an integration constant by imposing the boundary
condition $p(\rho)/b_2(\rho) \rightarrow 0$ for $\rho \rightarrow
\infty$.

Comparing \eqref{zz6DMP}, \eqref{bsols} and \eqref{psol} with
\eqref{flatrzz} for six-dimensional Minkowski space we see that we
have the right asymptotic behavior, as discussed above. One can
compute that $\lambda=1$ which again is guaranteed by
Eq.~\eqref{rderlamb} and by choosing the right asymptotics.

For the six-dimensional Myers-Perry black hole we find a domain
structure with three domains $D_1$, $D_2$ and $D_3$ with
corresponding directions
\begin{equation}
W_1 = \frac{\partial}{\partial \phi_1}  \spa W_2 =
\frac{\partial}{\partial \phi_2} \spa W_3 = \frac{\partial}{\partial
t} + \Omega_1 \frac{\partial}{\partial \phi_1} + \Omega_2
\frac{\partial}{\partial \phi_2}
\end{equation}
We see that while the two first directions correspond to the two
rotational Killing vector fields the third direction is instead the
null Killing vector of the event horizon with the angular velocities
given by $\Omega_i = a_i/(a_i^2 + r_h^2 )$. The three domains are
\begin{equation}
\begin{array}{c} \ds
D_1 = \Big\{ (z^1,z^2) \in \R^2 \Big| z^1 = b_1 (\rho) x, \ z^2 =
b_2(\rho) + p(\rho) x^2 , \ \rho\geq \rho_h, \ |x|\leq 1 \Big\}
\\[4mm] \ds
D_2 = \Big\{ (z^1,z^2) \in \R^2 \Big| z^1 = b_1 (\rho) y, \ z^2 =
b_2(\rho)(2y^2-1) + p(\rho) y^2 , \ \rho\geq \rho_h, \ |y|\leq 1
\Big\}
\\[4mm] \ds
D_3 = \Big\{ (z^1,z^2) \in \R^2 \Big| z^1 = b_1 (\rho_h) xy, \ z^2 =
b_2(\rho_h)(2y^2-1) + p(\rho_h)x^2 y^2 , \ |x|\leq 1, \ 0\leq y \leq
1 \Big\}
\end{array}
\end{equation}
This domain structure is depicted
in the bottom left diagram of Figure \ref{domplots}. We note that in terms of the $(\rho,\theta,\psi)$ coordinates the
three domains correspond to $D_1: \theta=0$, $D_2: \psi=0,\pi$ and
$D_3: \rho=\rho_h$.

\section{Domain structure of seven dimensional black holes}
\label{sec:sevendim}

In this section we analyze the known asymptotically flat
seven-dimensional exact solutions of the vacuum Einstein equations.
These are the Minkowski space, the Schwarzschild-Tangherlini black
hole and the Myers-Perry black hole. They all have four Killing
vector fields, which is the maximally possible number in seven
dimensions. In addition the Killing vector fields obey the condition
\eqref{ortcondition}. This means that the metrics can be put in the
canonical form \eqref{rzmetric} with $p=4$ and $n=3$.

\subsubsection*{Minkowski space}

The metric of seven-dimensional Minkowski space is
\begin{equation}
ds^2 = - dt^2 + d\rho^2  + \rho^2 ( \mu_1^2 d\phi_1^2 + \mu_2^2
d\phi_2^2 + \mu_3^2 d\phi_3^2 + d\theta^2 + \cos^2 \theta d\psi^2 )
\end{equation}
with the director cosines
\begin{equation}
\label{dircos7D} \mu_1 = \sin \theta \spa \mu_2 = \cos \theta \sin
\psi \spa \mu_3 = \cos \theta \cos \psi
\end{equation}
and with the coordinate ranges $0 \leq \theta,\psi \leq \pi/2$.
Using \eqref{rzmetric} we see that
\begin{equation}
\label{mink7r} r = \rho^3 \mu_1 \mu_2 \mu_3 = \rho^3 \sin \theta
\cos^2 \theta \sin \psi \cos \psi = \frac{1}{2} \rho^3 \sin \theta
\cos^2 \theta \sin 2\psi
\end{equation}
From this we get $e^{-4\nu} =  \rho^4 \cos^2 \theta [ 4\sin^2 \theta +
\cos^2 \theta \sin^2 (2\psi)]/4$.
In order to fit in the canonical form \eqref{rzmetric}
we need to find the $z^\alpha$ coordinates such that
$g_{rz^\alpha}=0$ and $\lambda=1$. We make the following ansatz
$z^\alpha = \rho^{k_\alpha} F_\alpha (\theta) ( \cos 2 \psi
)^{l_\alpha}$ with $\alpha=1,2$. Demanding that $g_{rz^\alpha}=0$ gives that the
functions $F_\alpha(\theta)$ are of the form $F_\alpha (\theta) = C_\alpha (\cos \theta)^{2l_\alpha}
(3\cos^2\theta - 2)^{\frac{k_\alpha}{2} - l_\alpha}$
where $C_\alpha$ are constants. One can furthermore infer that
$\lambda=1$ provided $4 C_1 C_2 = \pm 1/(l_1 k_2 - k_1 l_2)$, $k_1+k_2=4$ and $l_1+l_2=1$.
We choose therefore the $z^\alpha$ coordinates
\begin{equation}
\label{mink7z} z^1 = \frac{1}{2} \rho^2 \cos^2 \theta \cos 2\psi
\spa z^2 = \frac{1}{4} \rho^2 ( 3\cos^2 \theta -2)
\end{equation}

We now consider the domain structure of seven-dimensional Minkowski
space using the coordinates \eqref{mink7r} and \eqref{mink7z}. This
can be done by analyzing the coordinates $z^\alpha$ when $r=0$. We
find the domain structure
\begin{equation}
\label{7Dminkdomains}
\begin{array}{c} \ds
W_1 = \frac{\partial}{\partial \phi_1} \spa D_1 = \Big\{ (z^1,z^2)
\in \R^2 \Big| z^2 \geq \frac{1}{2} |z^1| \Big\} \\[4mm] \ds
W_2 = \frac{\partial}{\partial \phi_2} \spa D_2 = \Big\{ (z^1,z^2)
\in \R^2 \Big| z^1 \geq 0 , \  z^2 \leq \frac{1}{2} z^1 \Big\} \\[4mm] \ds
W_3 = \frac{\partial}{\partial \phi_3} \spa D_3 = \Big\{ (z^1,z^2)
\in \R^2 \Big| z^1 \leq 0 , \ z^2 \leq - \frac{1}{2} z^1 \Big\}
\end{array}
\end{equation}
We see that $D_1 \cup D_2 \cup D_3= \R^2$. This domain structure is
depicted in the top right diagram of Figure \ref{domplots}. We note
that in terms of the $(\rho,\theta,\psi)$ coordinates the three
domains correspond to $D_1: \theta=0$, $D_2: \psi=0$ and $D_3:
\psi=\pi/2$. The origin of Minkowski space $\rho=0$ is seen to be
the common intersection point of all of the three domains. This
makes sense since the origin is the only point which is a fixed
point of rotation in all of the three rotation planes.

Building on our parametrization of seven-dimensional Minkowski space
given by Eqs.~\eqref{mink7r} and \eqref{mink7z} we can now describe
the boundary conditions that we wish to impose on seven-dimensional
asymptotically flat space-times. We consider here solutions with
$p=3$ such that one can write them in the form \eqref{rzmetric2} in
terms of coordinates $(r,z^1,z^2)$. We define the asymptotic region
in $(r,z^1,z^2)$ coordinates as $L \rightarrow \infty$ with
$r^{2/3}/L$, $z^1/L$ and $z^2/L$ finite or going to zero where $L
\equiv r^{2/3} + |z_1| + |z_2|$. In this asymptotic region we
require that the metric should asymptote to seven-dimensional
Minkowski space. This means in particular that we require the
$(r,z^1,z^2)$ coordinates to asymptote to Eqs.~\eqref{mink7r} and
\eqref{mink7z}. For the domain structure at $r=0$ this means that
for $|z_1| + |z_2| \rightarrow \infty$ we have the three domains
\eqref{7Dminkdomains}, with the border between the domains being at
the curves $z^2 = |z^1| /2 $ and $z^1 = 0$ for $z^2 \leq 0$ up to
corrections of order $(|z_1| + |z_2|)^{-1}$.

\subsubsection*{Schwarzschild-Tangherlini black hole}

The 7D Schwarzschild-Tangherlini black hole has the metric
\begin{equation}
ds^2 = - f dt^2 + \frac{ d\rho^2}{f} + \rho^2 ( \mu_1^2 d\phi_1^2 +
\mu_2^2 d\phi_2^2 + \mu_3^2 d\phi_3^2   + d\theta^2 + \cos^2 \theta
d\psi^2 ) \spa f = 1 - \frac{\rho_0^4}{\rho^4}
\end{equation}
with the director cosines given by \eqref{dircos7D}. We get
\begin{equation}
\label{schw7r} r = \rho^3 \sqrt{f} \mu_1 \mu_2 \mu_3 = \rho^3
\sqrt{f} \sin \theta \cos^2 \theta \sin \psi \cos \psi = \frac{1}{2}
\rho^3 \sqrt{f} \sin \theta \cos^2 \theta \sin 2\psi
\end{equation}
From this one can easily compute $\exp(-4\nu)$ as function of
$(\rho,\theta,\psi)$. Make now the ansatz $ z^1 = b_1(\rho) \cos^2
\theta \cos 2\psi$ and $z^2 = b_2(\rho) (3\cos^2\theta -2)$.
Imposing $g_{rz^\alpha}=0$ is equivalent to $b_1'/b_1 = b_2'/b_2 =
6\rho^3 / (3\rho^4 - \rho_0^4)$. We get therefore the $z^\alpha$
coordinates
\begin{equation}
\label{schw7z} z^1 = \frac{1}{2} \sqrt{\rho^4 - \frac{\rho_0^4}{3} }
\cos^2 \theta \cos 2\psi \spa z^2 = \frac{1}{4} \sqrt{\rho^4 -
\frac{\rho_0^4}{3} } (3\cos^2\theta -2)
\end{equation}
Comparing this with \eqref{mink7r} and \eqref{mink7z} for
seven-dimensional Minkowski space we see that we have the right
asymptotic behavior, as discussed above. One can compute that
$\lambda=1$ which is guaranteed by Eq.~\eqref{rderlamb} and by
choosing the right asymptotics.

\begin{figure}[ht]
\centering
\includegraphics[height=11cm,width=15cm]{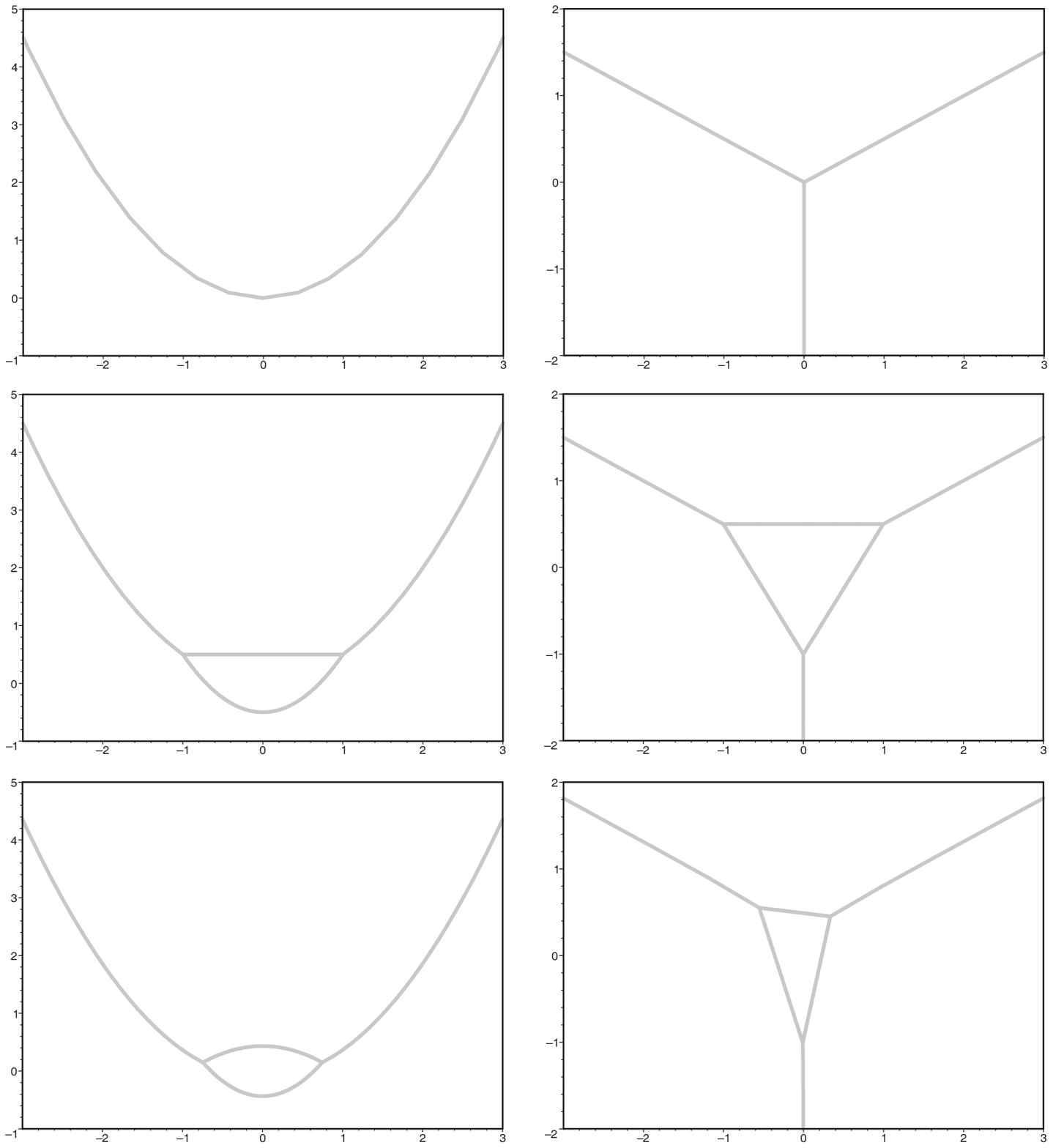}
\caption{{\small On the left side are shown the domain structures for the
six-dimensional Minkowski space (top left), the
Schwarzschild-Tangherlini black hole (middle left) with
$\rho_0^3=4/3$ and the Myers-Perry black hole (bottom left) with
$\rho_0^3=4/3$, $a_1=1/4$ and $a_2=4/5$.
On the right side are shown the domain structures for the
seven-dimensional Minkowski space (top right), the
Schwarzschild-Tangherlini black hole (middle right) with
$\rho_0^4=6$ and the Myers-Perry black hole (bottom right) with
$\rho_0^4=6$, $a_1=3/2$, $a_2=3/4$ and $a_3 = 1/3$.} \label{domplots}
}
\begin{picture}(442,0)(0,12)
\put(60,420){\footnotesize $W_1$}
\put(35,370){\footnotesize $W_2$}
\put(60,314){\footnotesize $W_1$}
\put(35,264){\footnotesize $W_2$}
\put(107,255){\footnotesize $W_3$}
\put(60,208){\footnotesize $W_1$}
\put(35,158){\footnotesize $W_2$}
\put(107,149){\footnotesize $W_3$}
\put(328,425){\footnotesize $W_1$}
\put(388,380){\footnotesize $W_2$}
\put(268,380){\footnotesize $W_3$}
\put(328,319){\footnotesize $W_1$}
\put(388,274){\footnotesize $W_2$}
\put(268,274){\footnotesize $W_3$}
\put(328,285){\footnotesize $W_4$}
\put(328,213){\footnotesize $W_1$}
\put(388,168){\footnotesize $W_2$}
\put(268,168){\footnotesize $W_3$}
\put(326,181){\footnotesize $W_4$}
\end{picture}
\end{figure}

The domain structure for the seven-dimensional
Schwarzschild-Tangherlini black hole as found from the coordinates
\eqref{schw7r} and \eqref{schw7z} is given by
\begin{equation}
\begin{array}{c} \ds
W_1 = \frac{\partial}{\partial \phi_1} \spa D_1 = \left\{ (z^1,z^2)
\in \R^2 \left| z^2 \geq \frac{1}{2} |z^1| , \ z^2 \geq \frac{\rho_0^2}{2\sqrt{6}} \right. \right\} \\[4mm] \ds
W_2 = \frac{\partial}{\partial \phi_2} \spa D_2 = \left\{ (z^1,z^2)
\in \R^2 \left| z^1 \geq 0 , \  z^2 \leq \frac{1}{2} z^1 , \ z^2 \leq \frac{3}{2} z^1 - \frac{\rho_0^2}{\sqrt{6}} \right. \right\} \\[4mm] \ds
W_3 = \frac{\partial}{\partial \phi_3} \spa D_3 = \left\{ (z^1,z^2)
\in \R^2 \left| z^1 \leq 0 , \ z^2 \leq - \frac{1}{2} z^1 , \ z^2 \leq - \frac{3}{2} z^1 - \frac{\rho_0^2}{\sqrt{6}} \right. \right\} \\[4mm] \ds
W_4 = \frac{\partial}{\partial t} \spa D_4 = \left\{ (z^1,z^2) \in
\R^2 \left| \frac{3}{2} |z^1| - \frac{\rho_0^2}{\sqrt{6}} \leq z^2
\leq \frac{\rho_0^2}{2\sqrt{6}} \right. \right\}
\end{array}
\end{equation}
This domain structure is depicted
in the middle right diagram of Figure \ref{domplots}. We note that in terms of the $(\rho,\theta,\psi)$ coordinates the
four domains correspond to $D_1: \theta=0$, $D_2: \psi=0$, $D_3:
\psi=\pi/2$ and $D_4: \rho=\rho_0$.

\subsubsection*{Myers-Perry black hole}

The seven-dimensional Myers-Perry black hole solution is \cite{Myers:1986un}
\begin{equation}
ds^2 = - dt^2 + \sum_{i=1}^3 (\rho^2 + a_i^2) ( d\mu_i^2 + \mu_i^2
d\phi_i^2 ) + \frac{\rho_0^{D-3} \rho^2 }{\Pi F} \Big( dt -
\sum_{i=1}^3 a_i \mu_i^2 d\phi_i \Big)^2 + \frac{\Pi F d\rho^2}{\Pi
- \rho^2 \rho_0^{4} }
\end{equation}
with the director cosines given by \eqref{dircos7D} and we have
\begin{equation}
F ( \rho , \mu_i ) = 1 - \sum_{i=1}^3 \frac{a_i^2 \mu_i^2}{\rho^2 +
a_i^2 } \spa \Pi(\rho) = \prod_{i=1}^3 ( \rho^2 + a_i^2 )
\end{equation}
The horizon is placed at $\rho=\rho_h$ which is defined as the
largest real root of the equation $\Pi (\rho) = \rho^2 \rho_0^4$.
From \eqref{rzmetric} we find
\begin{equation}
\label{mp7r} r = \sqrt{\Pi - \rho^2 \rho_0^4}\, \mu_1 \mu_2 \mu_3 =
\frac{1}{2} \sqrt{\Pi - \rho^2 \rho_0^4}\, \sin \theta \cos^2 \theta
\sin 2\psi
\end{equation}
We use the following ansatz for $z^\alpha$
\begin{equation}
\label{mp7z} z^\alpha = z_0^\alpha +  p_\alpha(\rho) \cos^2 \theta
\cos 2\psi + q_\alpha(\rho) (3\cos^2 \theta -2)
\end{equation}
The orthogonality conditions $g_{rz^\alpha}=0$ are equivalent to the
relations
\begin{equation}
\label{pversusq} \begin{array}{c} \rho(a_2^2-a_3^2) p_\alpha =  -
3\rho ( 2\rho^2+a_2^2+a_3^2) q_\alpha + ( 3\rho^4 + 2\rho^2 a^2 +
B^4 - \rho_0^4 ) q_\alpha' \\[4mm] \ds
3\rho(a_2^2-a_3^2) q_\alpha =  - \rho ( 6\rho^2+4a_1^2 +
a_2^2+a_3^2) p_\alpha + ( 3\rho^4 + 2\rho^2 a^2 + B^4 - \rho_0^4 )
p_\alpha'
\end{array}
\end{equation}
where we defined for convenience $a^2 \equiv a_1^2+a_2^2+a_3^2$, $B^4 \equiv a_1^2a_2^2 + a_1^2
a_3^2 + a_2^2 a_3^2$ and $C^4 \equiv a_1^4+a_2^4+a_3^4 - B^4$.
From these relations one can infer that $p_\alpha$ and $q_\alpha$
solve the same second order ODE which has the two independent
solutions
\begin{equation}
F_\pm (\rho) \equiv \sqrt{3 \rho^4 + 2 a^2 \rho^2 + B^4 - \rho_0^4}
\exp \left\{ \pm \frac{C^2}{\sqrt{3\rho_0^4 + C^4}} \,
\mbox{arctanh} \! \left( \frac{\sqrt{3\rho_0^4 + C^4}}{3\rho^2+a^2}
\right) \right\}
\end{equation}
Write now
\begin{equation}
p_\alpha(\rho) = p^+_\alpha F_+(\rho) + p^-_\alpha F_-(\rho) \spa
q_\alpha (\rho) = q^+_\alpha F_+(\rho) + q^-_\alpha F_-(\rho)
\end{equation}
One set of constraints on $p^\pm_\alpha$ and $q^\pm_\alpha$ comes
from demanding that $z^1$ and $z^2$ asymptotes to \eqref{mink7z} for
$\rho\rightarrow\infty$. This can be worked out using that
$F_\pm(\rho) \simeq \sqrt{3} \rho^2$ for $\rho \rightarrow \infty$.
Another set of constraints is that the equations \eqref{pversusq}
should be satisfied. This fixes
\begin{equation}
p_1^\pm = 2 q_2^\mp = \mp \frac{2a_1^2-a_2^2-a_3^2 \mp 2C^2}{8\sqrt{3} C^2} \spa
q_1^\pm = \frac{2}{3} p_2^\pm = \mp \frac{a_2^2-a_3^2}{8\sqrt{3} C^2}
\end{equation}
We furthermore impose that $z^1 \rightarrow 0$ for $\rho\rightarrow \infty$ when
$\theta = \pi/2 $  and $z^2|_{\psi=0} + z^2|_{\psi=\pi/2} \rightarrow 0$ for $\rho \rightarrow
\infty$ when $3\cos^2 \theta = 2 $. This fixes $z_0^1 = - (a_2^2-a_3^2)/6$ and $z_0^2 = 0$.

Comparing the coordinates \eqref{mp7r}-\eqref{mp7z} with those of
seven-dimensional Minkowski space \eqref{mink7r} and \eqref{mink7z}
we see that we have the right asymptotic behavior, as discussed
above. One can again compute that $\lambda=1$ which is guaranteed by
Eq.~\eqref{rderlamb} and by choosing the right asymptotics.

For the seven-dimensional Myers-Perry black hole we find a domain
structure with four domains $D_1$, $D_2$, $D_3$ and $D_4$ with
corresponding directions
\begin{equation}
W_1 = \frac{\partial}{\partial \phi_1}  \spa W_2 =
\frac{\partial}{\partial \phi_2} \spa W_3 = \frac{\partial}{\partial
\phi_3}\spa W_4 = \frac{\partial}{\partial t} + \Omega_1
\frac{\partial}{\partial \phi_1} + \Omega_2 \frac{\partial}{\partial
\phi_2} + \Omega_3 \frac{\partial}{\partial \phi_3}
\end{equation}
We see that while the three first directions correspond to the three
rotational Killing vector fields the fourth direction is instead the
null Killing vector of the event horizon with the angular velocities
given by $\Omega_i = a_i/(a_i^2+r_h^2)$. The four domains are
\begin{equation}
\begin{array}{c} \ds
D_1 = \Big\{ (z^1,z^2) \in \R^2 \Big| z^\alpha = z_0^\alpha +
p_\alpha(\rho) x + q_\alpha(\rho), \ \rho \geq \rho_h, \ |x|\leq 1
\Big\}
\\[4mm] \ds
D_2 = \Big\{ (z^1,z^2) \in \R^2 \Big| z^\alpha = z^\alpha_0 +
p_\alpha(\rho) y + q_\alpha(\rho) (3y-2), \ \rho\geq \rho_h , \
0\leq y \leq 1 \Big\}
\\[4mm] \ds
D_3 = \Big\{ (z^1,z^2) \in \R^2 \Big| z^\alpha = z^\alpha_0 -
p_\alpha(\rho) y + q_\alpha(\rho) (3y-2), \ \rho\geq \rho_h , \
0\leq y \leq 1 \Big\}
\\[4mm] \ds
D_4 = \Big\{ (z^1,z^2) \in \R^2 \Big| z^\alpha = z^\alpha_0 +
p_\alpha(\rho_h) yx + q_\alpha(\rho_h) (3y-2), \ |x| \leq 1 , \
0\leq y \leq 1 \Big\}
\end{array}
\end{equation}
This domain structure is depicted
in the bottom right diagram of Figure \ref{domplots}. We note that in terms of the $(\rho,\theta,\psi)$ coordinates the
four domains correspond to $D_1: \theta=0$, $D_2: \psi=0$, $D_3:
\psi=\pi/2$ and $D_4: \rho=\rho_h$.

\section{Possible new domain structures in six and seven dimensions}
\label{sec:possible}

In this section we examine the possible domain structures one can
have for asymptotically flat solutions in six and seven dimensions
with $D-3$ commuting linearly independent Killing vector fields.
We illustrate the domain structure diagrams in a different fashion
than in Sections \ref{sec:sixdim} and \ref{sec:sevendim} since here
we do not care about all details of the domain structure.

\subsection{Six-dimensional asymptotically flat space-times}
\label{sec:6Dposs}

We consider here the possible domain structures of asymptotically
flat six-dimensional black hole space-times with three commuting
linearly independent Killing vector fields.

\begin{figure}[ht]
\centering
\includegraphics[height=6.5cm,width=10cm]{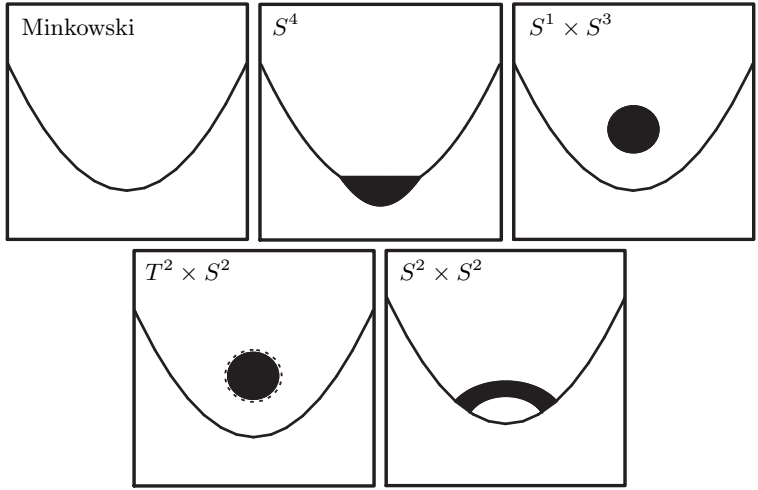}
\caption{{\small Domain structure for six-dimensional Minkowski
space and four possible domain structures for six-dimensional
asymptotically flat black holes with a single event horizon.}
\label{6Ddomains} }
\begin{picture}(430,0)(0,0)
\put(79,229){\footnotesize Minkowski} \put(174,229){\footnotesize
$S^4$ } \put(271,229){\footnotesize $S^1\times S^3$ }
\put(126,136){\footnotesize $T^2 \times S^2$ }
\put(222,136){\footnotesize $S^2 \times S^2$ }
\end{picture}
\end{figure}

In the first diagram of Figure \ref{6Ddomains} we have depicted the
domain structure of six-dimensional Minkowski space. Here the upper
domain has direction $\partial / \partial \phi_1$ and the lower
domain direction $\partial / \partial \phi_2$. These two domains
correspond to the set of fixed points of the rotations in two
rotation planes of six-dimensional Minkowski space. The idea is now
to examine all the possible ways in which we can put a domain with a
time-like direction corresponding to an event horizon in this domain
structure diagram. We represent the event horizon domain as a filled
area. Over this domain is fibred two circles parameterized by the
two rotation angles $\phi_1$ and $\phi_2$. The topology of the event
horizon is now determined from where these two circles shrink to
zero at the boundary of the domain. This give rise to three distinct
types of event horizons corresponding to $S^4$, $S^1 \times S^3$ or
$S^2\times S^2$ topology as we discuss below. Another possibility is
that the domain structure do not live in the plane $\R^2$ but in a
disconnected space. As we discuss below this can give rise to an
event horizon with $T^2 \times S^2$ topology.

The first possibility is to put the event horizon domain across the
boundary of the two rotational domains. This is depicted in the
second diagram of Figure \ref{6Ddomains} where we chose for
convenience the filled area to have a shape corresponding to an area
in between the branches of a parabola. We see that the boundary of
the domain is divided in two parts, one in which the first circle is
shrunk to zero, the other in which the second circle is shrunk to
zero. This corresponds to the topology of a four-sphere. This is
shown explicitly in Appendix \ref{sec:para}. From comparing with
Figure \ref{domplots} we see that this domain structure indeed is
equivalent to those of the six-dimensional Schwarzschild-Tangherlini
and Myers-Perry black hole.

The second possibility is to put the event horizon domain away from
the curve separating the two rotational domains. This gives two
possibilities, depending on whether we put it above or below.
However these two possibilities are equivalent by relabeling the two
rotation planes. We have illustrated one of the possibilities in the
third diagram of Figure \ref{6Ddomains}. In such a space-time the
event horizon is topologically an $S^1 \times S^3$. This is seen
from the fact that we again have two circles fibred over a disc but
on the boundary the one parameterized by $\phi_1$ shrinks to zero
while the other one is of non-zero size everywhere on the event
horizon. As shown explicitly in Appendix \ref{sec:para} a circle
fibred over a disc for which the circle shrinks to zero at the
boundary of the disc corresponds to a three-sphere topology. Thus,
the domain structure corresponds to a black ring in six dimensions.
Approximate metrics for neutral black rings in the ultraspinning
regime have been found in \cite{Emparan:2007wm} and described using
the Blackfold approach in \cite{Emparan:2009cs}.

The third possibility is that the event horizon domain is
disconnected from the rotational domains. This can happen if the
event horizon is displaced from the fixed points of rotations in
both of the rotation planes. In \cite{Emparan:2009cs} an example of
this called a black torus is described with $T^2 \times S^2$
topology using the Blackfold approach again in the ultra-spinning
regime. This is realized by having the domain submanifold $B = \R^2
\cup S^2$. This is concretely realized as having the domain plane
parameterized by $(z^1,z^2)$ being multi-valued so that for the
$(z^1,z^2)$ values where we have the event horizon domain we have
three sheets of the domain plane --~one sheet corresponding to the
domain structure of the six-dimensional Minkowski space and the two
other sheets disconnected from this being the two sides of a
two-sphere projected on to a plane, see Appendix \ref{sec:para} for
an explicit parametrization of this. In the fourth diagram of Figure
\ref{6Ddomains} we have depicted this domain structure where the
dashed line represents that the event horizon domain is disconnected
from the two rotational domains. Clearly a space-time with such a
domain structure has an event horizon with $T^2 \times S^2$
topology, with $T^2 = S^1 \times S^1 $ being a rectangular torus,
since the two circles do not shrink to zero at any point on the
event horizon domain.

Finally, the fourth possibility for a domain structure is depicted
in the fifth diagram of Figure \ref{6Ddomains}. We see that the
event horizon here is shaped as a piece of a ring. The event horizon
can be seen to have an $S^2\times S^2$ topology since in the angular
direction we have that the $\phi_2$ circle shrinks to zero in the
two ends, while in the radial direction the $\phi_1$ circle shrinks
to zero in the two ends. Unlike the three above domain structures we
do not have any evidence that this domain structure corresponds to a
regular black hole space-time. However, numerical evidence for a
static black hole space-time with this domain structure, though with
a conical singularity, has been found in \cite{Kleihaus:2009wh}. In
the Blackfold approach \cite{Emparan:2009cs} this kind of event
horizon topology has also been considered in the limit in which one
sphere is much larger than the other. It was found that the sphere
cannot be supported by a single large angular momentum in this
limit. However, it is conceivable that the solution can be made
regular by having two angular momenta turned on, one for each
sphere.

\subsubsection*{Multiple horizons}

\begin{figure}[ht]
\centering
\includegraphics[height=3cm,width=10cm]{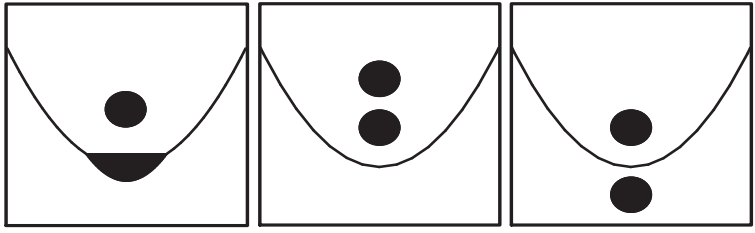}
\caption{{\small Three possible domain structures for
six-dimensional asymptotically flat black holes with two separate
event horizons.} \label{6Dmultiple} }
\end{figure}

It is interesting to consider the combinations one can make of the
above domain structures. For simplicity we restrict ourselves to the
first two possibilities depicted in Figure \ref{6Ddomains}. First we
can make a Black Saturn, i.e. a black ring with a black hole in the
center. This corresponds to the domain structure depicted in the
first diagram of Figure \ref{6Dmultiple}. In five dimensions such a
solution has been found in \cite{Elvang:2007rd}. We can also make a
di-ring, i.e. two rings which are concentric and rotating in the
same rotation plane. This corresponds to the domain structure of the
second diagram of Figure \ref{6Dmultiple}. In five dimensions such a
solution has been found in \cite{Iguchi:2007is}. Finally we can
imagine two bicycling black rings. These rotate in two orthogonal
rotation planes. This corresponds to the domain structure of the
third diagram of Figure \ref{6Dmultiple}. In five dimensions such a
solution has been found in \cite{Izumi:2007qx}.

\subsection{Seven-dimensional asymptotically flat space-times}
\label{sec:7Dposs}

We consider here the possible domain structures of asymptotically
flat seven-dimensional black hole space-times with four commuting
linearly independent Killing vector fields.

\begin{figure}[ht]
\centering
\includegraphics[height=7cm,width=12cm]{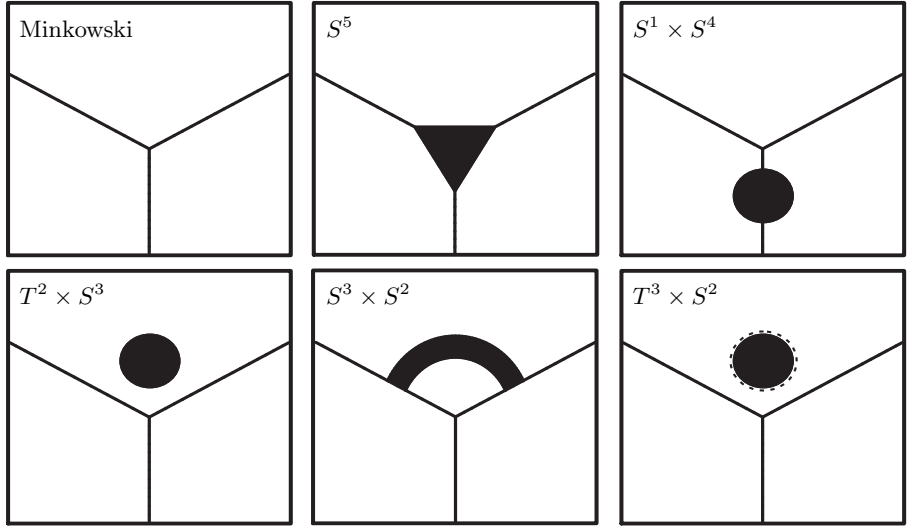}
\caption{{\small Domain structure for seven-dimensional Minkowski
space and five possible domain structures for seven-dimensional
asymptotically flat black holes with a single event horizon.}
\label{7Ddomains}}
\begin{picture}(430,0)(0,0)
\put(50,242){\footnotesize Minkowski} \put(166,242){\footnotesize
$S^5$ } \put(282,242){\footnotesize $S^1\times S^4$ }
\put(50,141){\footnotesize $T^2 \times S^3$ }
\put(166,141){\footnotesize $S^3\times S^2$ }
\put(282,141){\footnotesize $T^3 \times S^2$ }
\end{picture}
\end{figure}

In the first diagram of Figure \ref{7Ddomains} we have depicted
seven-dimensional Minkowski space. Here the upper domain has
direction $\partial / \partial \phi_1$, the right domain has
direction $\partial / \partial \phi_2$ and the left domain has
direction $\partial / \partial \phi_3$. These three domains
correspond to the set of fixed points of the rotations in three
rotation planes of seven-dimensional Minkowski space. We now want to
examine all the possible ways in which we can put a domain with a
time-like direction corresponding to an event horizon in this
diagram. We represent this domain as a filled area. Over this domain
is fibred three circles parameterized by the three rotation angles
$\phi_1$, $\phi_2$ and $\phi_3$. The topology of the event horizon
is now determined from where these three circles shrink to zero at
the boundary of the domain. This give rise to four distinct types of
event horizons corresponding to the topologies $S^5$, $S^1 \times
S^4$, $T^2\times S^3$ and $S^3\times S^2$ as we discuss below. It is
furthermore possible to draw a domain structure that gives rise to a
topology with identifications of the five-sphere as we describe
below. Another possibility is that the domain structure does not
live in the plane $\R^2$ but in a disconnected space. As we discuss
below this can give rise to an event horizon with $T^3\times S^2$
topology.

The first possibility is that the event horizon domain covers the
origin of the seven-dimensional Minkowski space - i.e. the point
belonging to all three rotational domains. This is depicted in the
second diagram of Figure \ref{7Ddomains} where the filled area for
convenience has the shape of a triangle. We see that the boundary of
the domain is divided in three parts, one for each side of the
triangle. At each side of the triangle a different circle shrinks to
zero. From this one can infer that the event horizon has topology of
a five-sphere. This is shown explicitly in Appendix \ref{sec:para}.
Comparing this domain structure with Figure \ref{domplots} we see
that it is equivalent to those of the seven-dimensional
Schwarzschild-Tangherlini black hole and Myers-Perry black hole.

The second possibility is that the event horizon domain crosses one
of the curves dividing the three rotational domains. This domain
structure is depicted in the third diagram of Figure
\ref{7Ddomains}. It gives rise to an event horizon with $S^1 \times
S^4$ topology where the $S^1$ corresponds to the $\phi_1$ circle
since that is finite everywhere on the event horizon domain. Instead
with respect to the $\phi_2$ and $\phi_3$ circles we see that the
boundary of the event horizon domain is divided in two parts, one
part where the $\phi_2$ circle shrinks to zero, the other where the
$\phi_3$ shrinks to zero. As shown in Section \ref{sec:6Dposs} this
corresponds to an $S^4$ topology. Thus, this domain structure
corresponds to a black ring in seven dimensions. Approximate metrics
for neutral black rings in the ultraspinning regime have been found
in \cite{Emparan:2007wm} and described using the Blackfold approach
in \cite{Emparan:2009cs}.

The third possibility is that the event horizon domain do not cross
any of the curves dividing the three rotational domains. This domain
structure is depicted in the fourth diagram of Figure
\ref{7Ddomains}. This corresponds to an event horizon with $T^2
\times S^3$ topology where the rectangular torus $T^2 = S^1 \times
S^1$ corresponds to the $\phi_2$ and $\phi_3$ circles since they do
not shrink to zero on the event horizon domain. Instead the $\phi_1$
circle shrinks to zero on the boundary of the event horizon domain
hence this gives rise to the $S^3$ part of the topology, as shown in
Section \ref{sec:6Dposs}. Thus, this domain structure corresponds to
a so-called black torus which has been described using the Blackfold
approach in \cite{Emparan:2009cs}.

The fourth possibility is that the event horizon domain covers an
area in between two of the curves dividing the three rotational
domains. This domain structure is depicted in the fifth diagram of
Figure \ref{7Ddomains} with the shape of a piece of a ring. This
corresponds to an event horizon with $S^3 \times S^2$ topology. This
is seen from the fact that the boundary of the domain is split up in
four intervals, each corresponding to a side of the domain,
according to which circle shrinks to zero. For the upper and lower
sides the $\phi_1$ circle shrinks to zero while for the left and
right sides either the $\phi_2$ or $\phi_3$ circle shrinks to zero.
This clearly gives an $S^3\times S^2$ topology since when we go from
boundary to boundary in the angular direction we go from shrinking
the $\phi_2$ circle to shrinking the $\phi_3$ circle, thus giving a
three-sphere, and when we go from boundary to boundary in the radial
direction we shrink the $\phi_1$ circle at both boundaries thus
giving a two-sphere. In the Blackfold approach of
\cite{Emparan:2009cs} such an event horizon topology has been found
in the limit where the $S^2$ is much smaller than the $S^3$,
corresponding to the limit where the upper and lower sides in the
fifth diagram of Figure \ref{7Ddomains} are very close.

The domain structure giving a $S^3\times S^2$ topology is of
particular interest since we see that it has a finite size domain
with a space-like direction. This is very reminiscent of the rod
structure of the five-dimensional black ring where one has a finite
space-like rod. In particular this means that the domain structure
has two invariants corresponding to the areas of the two domains of
finite size, as measured using the metric \eqref{Bmet}.

The finite size space-like domain also provide a possible generalization. If we let the direction of this finite size domain be
\begin{equation}
W = \frac{\partial}{\partial \phi_1} + q  \frac{\partial}{\partial \phi_2}
\end{equation}
with $q$ an integer, then we see that we have a Lens space
$L(q,1)=S^3/\Z_q$ when going in the radial direction. Instead in the
angular direction we still have an $S^3$ in terms of the $\phi_2$
and $\phi_3$ circles. Thus, the topology of the event horizon is now
$S^5 / \Z_q$, which is a five-dimensional Lens space. This is
reminiscent of what happens for five-dimensional black holes where
one can get a three-dimensional Lens space by changing the direction
of the finite space-like rod in the rod structure of the black ring
\cite{Hollands:2007aj,Evslin:2008gx}.

Finally, the last possibility considered here is that the event
horizon domain is disconnected from the rotational domains. Thus the
event horizon is displaced from the fixed points of rotation in all
three rotation planes. This works the same way as in six dimensions.
The domain submanifold is $B = \R^2 \cup S^2$ and it can again be
viewed as a three-sheeted plane. This gives a three-torus topology
$T^3 \times S^2$, with $T^3 = S^1\times S^1\times S^1$ a rectangular
three-torus, and such a black three-torus have indeed been described
by the Blackfold approach in \cite{Emparan:2009cs}. We depicted the
domain structure in the sixth diagram of Figure \ref{7Ddomains}.

\subsubsection*{Multiple horizons}

\begin{figure}[ht]
\centering
\includegraphics[height=3cm,width=14cm]{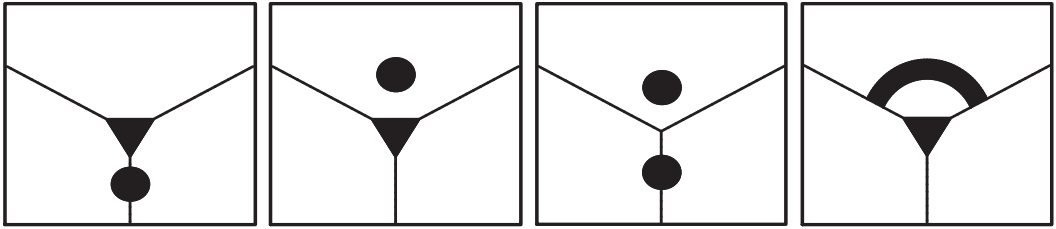}
\caption{{\small Four possible domain structures for
seven-dimensional asymptotically flat black holes with two separate
event horizons.} \label{7Dmultiple}}
\end{figure}

Just as in six dimensions it is again interesting to consider the
combinations one can make of the above examples of domain structures
for seven-dimensional black holes. Examples of this include the
seven-dimensional version of the Black Saturn, see first diagram of
Figure \ref{7Dmultiple}, a black hole ($S^5$ topology) with a black
torus ($T^2\times S^3$ topology) around, see second diagram of
Figure \ref{7Dmultiple}, a black ring ($S^1\times S^4$ topology)
with a black torus ($T^2\times S^3$ topology) around, see third
diagram of Figure \ref{7Dmultiple}, and an black hole ($S^5$
topology) with a black three-sphere around ($S^3\times S^2$
topology), see fourth diagram of Figure \ref{7Dmultiple}.

\section{Discussion and outlook}
\label{sec:concl}

In this paper we have introduced the domain structure for black hole space-times. We have shown that the domain structure provides invariants for a given space-time and that these invariants therefore can be part of the characterization of the space-time.
A natural question following this is whether these invariants are enough to give a complete characterization of a black hole space-time.

We first restrict ourselves to solutions of the vacuum Einstein
equations, and assume furthermore that the orthogonality condition
\eqref{ortcondition} is obeyed. For stationary and asymptotically
flat solutions with $[(D-1)/2]$ rotational Killing vector fields we
have the highest number of Killing vector fields possible for
solutions which are not Minkowski space.
It is natural to make the following conjecture:%
\footnote{Note that we added an assumption on connectedness of $B$
since it is unclear whether the domain structure contains enough
information to parameterize a situation with disconnected $B$.}

\begin{conjecture}
\label{uniqconj} Let two $D$-dimensional regular and stationary
asymptotically flat solutions of the vacuum Einstein equations be
given, both with a single connected event horizon and with
$[(D-1)/2]$ commuting rotational Killing vector fields obeying the
orthogonality condition \eqref{ortcondition}. Let the two solutions
have the same mass and angular momenta. Assume that the set $B$ is
connected for both solutions. Then the two solutions are the same if
and only if they have the same domain structure. \squ
\end{conjecture}

For $D=5$ this is shown to be true \cite{Hollands:2007aj} following
the uniqueness hypothesis of \cite{Harmark:2004rm}. However, for
$D>5$ it is clear that one cannot apply the techniques used for
$D=4,5$. The problem is that the metric $\tilde{g}_{ab}$ on $\CN_n$
is not decoupled from the Killing vector metric $G_{ij}$ in the
Einstein equations. Thus, when given two solutions with the same
domain structure they will, generically, have both two different
$G_{ij}$ metrics as well as two different $\tilde{g}_{ab}$ metrics.
This means proving a uniqueness theorem is a considerably more
challenging task than for $D=5$ where it was enough to generalize
the methods introduced for $D=4$
\cite{Morisawa:2004tc,Hollands:2007aj}.

Consider now the general case, $i.e.$ without the orthogonality
condition \eqref{ortcondition} or restrictions on what matter fields
are present. Here we observe that the domain structure in general is
not enough to fully characterize a solutions. To see this we can use
a lesson from the paper \cite{Emparan:2004wy} where a ring with a
dipole charge was found. This gives infinite non-uniqueness of the
solution since the ring carries no net charge, as measured at
infinity. Clearly the domain structure cannot carry information on
the dipole charge thus it is evident that one needs to supplement
the domain structure invariants with information about locally
measured charges, such as the dipole charge (see
\cite{Copsey:2005se} for a general exposition on dipole charges and
other local charges). It would be very interesting to pursue this
problem further to find a general way to specify the dipole charges
- as well as other types of local charges - for the event horizon
domains. Combined with the domain structure this could lead to a
full characterization of asymptotically flat black hole space-times
with $[(D-1)/2]$ rotational Killing vector fields.

Another direction which is interesting to consider is asymptotically
flat solutions with less than $[(D-1)/2]$ rotational Killing vector
fields. As an example we can take the case of $D=5$. Consider a
stationary, but non-static, black hole. Write the null Killing
vector of the event horizon as
\begin{equation}
W = \frac{\partial}{\partial t} + \Omega_1 \frac{\partial}{\partial
\phi_1} + \Omega_2 \frac{\partial}{\partial \phi_2}
\end{equation}
Then we know from the Rigidity theorems of \cite{Hollands:2006rj}
that $W$ is a Killing vector field of the space-time. Thus, the
space-time have the two Killing vector fields
\begin{equation}
V_{(0)} = \frac{\partial}{\partial t} \spa V_{(1)} =
p\frac{\partial}{\partial \phi_1} + q \frac{\partial}{\partial
\phi_2}
\end{equation}
with $\Omega_1/\Omega_2=p/q$. We observe now that we can assume $p$
and $q$ to be relatively prime numbers since $V_{(1)}$ should
generate a $U(1)$. In other words $\Omega_1/\Omega_2$ is a rational
number. One can now proceed with finding the domain structure of the
solution.

Another interesting direction to pursue is to consider the various
possible domain structures of solutions with black holes attached to
Kaluza-Klein bubbles for space-times which are asymptotically
Kaluza-Klein space $\R^{1,D-1-q} \times T^q$. As explored via the
rod structure in \cite{Elvang:2004iz} for asymptotically $\R^{1,4}
\times S^1$ and $\R^{1,5} \times S^1$ space-times solving vacuum
Einstein equations, this could lead to interesting new possibilities
for event horizon topologies.

We remark that the vacuum Einstein equations for solutions with
$D-3$ Killing vector fields have an enhanced symmetry, following the
construction \cite{Maison:1979kx}. The vacuum Einstein equations can
be written as a three-dimensional sigma-model with the target space
being an $SL(D-2,\R)$ group manifold. This is relevant for
asymptotically flat solutions of vacuum Einstein equations in six
and seven dimensions. It would be interesting to understand if one
can find similar hidden symmetries in the Einstein equations for
less number of Killing vectors. This could potentially lead to
algebraic solution generating techniques for $D>5$ similar to the
one proposed in \cite{Giusto:2007fx} for $D=5$.

Finally, as remarked previously, the domain structure can be
generalized to black hole space-times which are not asymptotically
flat, including asymptotically Anti-de Sitter space-times. This will
be considered in a future publication \cite{Harmark:domain2}.

\section*{Acknowledgments}

We thank Pau Figueras for useful discussions. We thank the Carlsberg
foundation for support. We thank the Galileo Galilei Institute at
Firenze, Italy, the Summer Institute 2008 at Mt.~Fuji, Japan, the
black hole workshop at Veli Losinj, Croatia, the CERN TH Institute
program on black holes, the Banff International Research Station,
Canada, and Perugia University at Perugia, Italy, for warm
hospitality while this project was carried out.

\begin{appendix}

\section{Einstein equations for space-times with Killing vector fields}
\label{sec:einstein}

In this appendix we give first a general expression for the Ricci
tensor for $D$-dimensional space-times with $p$ commuting Killing
vector fields. We then use this to write down the vacuum Einstein
equations for $D$-dimensional space-times with $p$ commuting Killing
vector fields. Finally we examine under which conditions the metric
can be written in a block diagonal form with the Killing part of the
metric being orthogonal to the rest of the metric.

\subsubsection*{Ricci tensor for space-times with Killing vector
fields}

We consider here a given $D$-dimensional manifold $\CM_D$ with a
metric with $p$ commuting linearly independent Killing vector fields
$V_{(i)}$, $i=0,1,...,p-1$. Define $n= D-p$. We can always find a
coordinate system $x^0,x^1,...,x^{p-1},y^1,...,y^n$ such that in
this coordinate system the Killing vectors are of the form
\eqref{killv} and the metric is of the form \eqref{genmet}, where
$G_{ij}$, $A^i_a$ and $\tilde{g}_{ab}$ only depend on $y^a$. Define
$K^2 = |\det G_{ij} |$, $\tilde{g} = | \det \tilde{g}_{ab} |$ and
$F^i_{ab} =
\partial_a A^i_b - \partial_b A^i_a$.
The components of the Ricci tensor are
\begin{align}
\label{ricci_ij} R_{ij} = & - \frac{1}{2} \partial_a \partial^a
G_{ij} - \frac{1}{2}
\partial_a ( \log K + \log \sqrt{\tilde{g}} ) \partial^a G_{ij} +
\frac{1}{2} \partial^a G_{ik} G^{kl} \partial_b G_{lj} \nn \\ & + \frac{1}{4} G_{ik} G_{jl} \tilde{g}^{ac} \tilde{g}^{bd} F^k_{ab}
F^l_{cd}
\\
R_{ia} = & {}\, R_{ij} A_a^j + \frac{1}{2K \sqrt{\tilde{g}}} \tilde{g}_{ab}
\partial_c \big( K \sqrt{\tilde{g}} G_{ij} \tilde{g}^{bd} \tilde{g}^{ce} F^j_{de} \big)
\\
R_{ab} = & - R_{ij} A^i_a A^j_b + R_{ia} A^i_b + R_{ib} A^i_a - \frac{1}{2} \tilde{g}^{cd} G_{ij} F^i_{ac} F^j_{bd} \nn \\ & +  \tilde{R}_{ab} - \tilde{D}_a \tilde{D}_b \log K - \frac{1}{4} \tr ( G^{-1}
\partial_a G G^{-1} \partial_b G )
\end{align}
with $\tilde{D}_a \tilde{D}_b \log K = \partial_a \partial_b \log K -
\tilde{\Gamma}^c_{ab} \partial_c \log K$
where $\tilde{\Gamma}^c_{ab}$ is the Christoffel symbol as computed
from the $\tilde{g}_{ab}$ metric.
Define
${(C_a)^i}_j = G^{ik} \partial_a G_{kj}$.
We can write the vacuum Einstein equations $R_{\mu\nu}=0$
as
\begin{eqnarray}
\label{ein1}
& \ds \partial_a ( K \sqrt{\tilde{g}}\, \tilde{g}^{ab} {(C_b)^i}_j )  = \frac{1}{2} K \sqrt{\tilde{g}}\, F^i_{ab} G_{jk} \tilde{g}^{ac} \tilde{g}^{bd} F^k_{cd} &
\\
\label{ein2} & \ds
\partial_a ( K \sqrt{\tilde{g}} G_{ij} \tilde{g}^{ac} \tilde{g}^{bd} F^j_{cd} ) =0 &
\\ & \ds
\label{ein3} \tilde{R}_{ab} = \frac{1}{4} \tr ( C_a C_b ) +
\tilde{D}_a \tilde{D}_b \log K + \frac{1}{2} \tilde{g}^{cd} G_{ij}
F^i_{ac} F^j_{bd} &
\end{eqnarray}
%

\subsubsection*{Results on mixed part of metric solving vacuum
Einstein equations}

We examine in this section under which conditions one can turn off
the $A^i_a$ fields in the general expression for a metric
\eqref{genmet} solving the vacuum Einstein equations \eqref{ein1}-\eqref{ein3}. The $A^i_a$ field corresponds to the mixed part of the metric \eqref{genmet} having indices both in the $x^i$ and $y^a$ directions.

\begin{theorem}
\label{ortform} Consider a solution of the vacuum Einstein equations
with $p$ commuting Killing vector fields $V_{(i)}$, $i=0,1,...,p-1$.
If the tensors $V_{(0)}^{[\mu_1} V_{(1)}^{\mu_2} \cdots
V_{(p-1)}^{\mu_{p}} D^\nu V_{(i)}^{\rho]}=0$ for all $i=0,1,...,p-1$
then we can find coordinates such that the metric is of
the form
\begin{equation}
\label{lgenmet} ds^2 = G_{ij} dx^i dx^j + \tilde{g}_{ab} dy^a dy^b
\end{equation}
with the Killing vector fields given by Eq.~\eqref{killv}.

\proof Define the one-forms $\xi^{(i)}_\mu = g_{\mu\nu} V_{(i)}^\nu$
for $i=0,1,...,p-1$. These one-forms span a $p$-dimensional linear
space $T^*$. Since $V_{(0)}^{[\mu_1} V_{(1)}^{\mu_2} \cdots
V_{(p-1)}^{\mu_{p}} D^\nu V_{(i)}^{\rho]}=0$ we see that $\xi^{(0)}
\wedge \xi^{(1)} \wedge \cdots \wedge d \xi^{(i)} = 0$ for all
$i=0,1,...,p-1$. This means that for any $\xi \in T^*$ we can find
one-forms $\psi^{(i)}$, $i=0,1,...,p-1$, such that $d\xi = \sum_{i=0}^{p-1} \psi^{(i)} \wedge \xi^{(i)}$.
Consider now the $n=D-p$ dimensional tangent space at each point
defined by being orthogonal to all one-forms in $T^*$. From
Frobenius' theorem we get that this collection of tangent spaces
admits integrable $n$-dimensional submanifolds. Hence we can find
coordinates such that the metric is of the form \eqref{lgenmet}.
\squ
\end{theorem}

To get another perspective on Theorem \ref{ortform} we introduce for a given solution
of the vacuum Einstein equations with $p$ commuting Killing vector
fields $V_{(i)}$ the $(n-2)$-forms $B_i$, $i=0,1,...,p-1$, as
\begin{equation}
\label{defbs} (B_i)_{\mu_1 \cdots \mu_{n-2} } = \sqrt{g} \epsilon_{\mu_1 \cdots
\mu_{n-2} \nu_1 \cdots \nu_{p} \rho \sigma} V_{(0)}^{\nu_1}
V_{(1)}^{\nu_2} \cdots V_{(p-1)}^{\mu_{p}} D^{\rho} V_{(i)}^{\sigma}
\end{equation}
where the $\epsilon$ is the $D$-dimensional $\epsilon$ symbol and
$g$ is the numerical value of the determinant of the $D$-dimensional
metric. In the $(x^i,y^a)$ coordinates of Eq.~\eqref{genmet} we see
that the $\mu_j$ indices only can take values in the $y^a$
directions. We compute now $(B_i)_{a_1 \cdots a_{n-2} } =
\frac{1}{2}  K \sqrt{\tilde{g}}\, \epsilon_{a_1 \cdots a_{n-2} bc}
\tilde{g}^{bd} \tilde{g}^{ce} G_{ij} F^j_{de}$ where the $\epsilon$
is the $n$-dimensional $\epsilon$ symbol and where $a_j,b,c =
1,...,n$. From this we see that $V_{(0)}^{[\mu_1} V_{(1)}^{\mu_2}
\cdots V_{(p-1)}^{\mu_{p}} D^\nu V_{(i)}^{\rho]}=0$ if and only if
$F^i_{ab}=0$. Therefore Theorem \ref{ortform} tells us that if
$F^i_{ab}=0$ then we can find a gauge transformation such that
$A^i_a=0$. This is already clear locally but Frobenius' theorem
ensures that it is also true globally. Another important property of
the $(n-2)$-forms $B_i$ is the following, which follows from the
above and Eq.~\eqref{ein2}.

\begin{lemma}
\label{lemmadb} Consider a solution of the vacuum Einstein equations
with $p$ commuting Killing vector fields $V_{(i)}$, $i=0,1,...,p-1$.
Then the $(n-2)$-forms defined in \eqref{defbs} are closed $dB_i=0$.
 \squ
\end{lemma}

Using this lemma we can prove the following theorem

\begin{theorem}
\label{usualcase} Consider a solution of the vacuum Einstein
equations with $D-2$ commuting Killing vector fields $V_{(i)}$,
$i=0,1,...,D-3$. If the tensor $V_{(0)}^{[\mu_1} V_{(1)}^{\mu_2}
\cdots V_{(D-3)}^{\mu_{D-2}} D^\nu V_{(i)}^{\rho]}=0$ vanishes at at
least one point of the manifold for any given $i=0,1,2,...,D-3$ then
we can write the metric of the solution in the form \eqref{lgenmet}
with the Killing vector fields given by \eqref{killv}.

\proof This follows from Theorem \ref{ortform} and Lemma
\ref{lemmadb} since in this case $B_i$ are scalar fields and hence
it follows from $dB_i=0$ for any given $i=1,2,...,D-2$ that
$B_i$ is constant on the manifold. \squ
\end{theorem}

This theorem is due to Wald in his book \cite{Wald:1984} and has
been generalized to any dimension by Emparan and Reall in
\cite{Emparan:2001wk}. Thus for $p=D-2$ it is enough that
$V_{(0)}^{[\mu_1} V_{(1)}^{\mu_2} \cdots V_{(D-3)}^{\mu_{D-2}} D^\nu
V_{(i)}^{\rho]}$ vanishes at single points for getting the form
\eqref{lgenmet} of the metric. Instead for $p < D-2$ we cannot write
a generic solution of the vacuum Einstein equations with $p$
commuting Killing vector fields in the form \eqref{lgenmet}.

\section{Parameterizations of topologies from domain structure}
\label{sec:para}

We consider the topologies that one can infer from a number of
circles fibred over a domain such that the circles shrinks to zero
at different points on the boundary of the domain.

 {\bf Four-sphere topology:}
We consider here two circles parameterized by $\phi_{1}$ and
$\phi_2$ fibred over a domain with the shape of the area between the
two branches of a parabola taken here to be $z^2 = (z^1)^2$ and
furthermore $z^2 \leq 1$. Write the embedding of a four-sphere as
\begin{equation}
x^1 + i x^2 = \sin \theta e^{i\phi_1} \spa x^3 + i x^4 = \cos \theta
\sin \psi e^{i\phi_2} \spa x^5 = \cos \theta \cos \psi
\end{equation}
where $0 \leq \theta \leq \pi/2$ and $0 \leq \psi \leq \pi$. We then
parameterize the domain as
\begin{equation}
z^1 = \cos \theta \cos \psi \spa z^2 = \cos^2 \theta
\end{equation}
We see that the $\phi_1$ circle shrinks to zero for the part of the
boundary where $z^2 = 1$ while the $\phi_2$ circle shrinks to zero
for the part where $z^2=(z^1)^2$.

 {\bf Three-sphere topology:}
We consider here a circle parameterized by $\phi_1$ fibred over a
domain with the shape of a disc $(z^1)^2 + (z^2 )^2 \leq 1$. Write
the embedding of a three-sphere as
\begin{equation}
x^1 + i x^2 = \cos \theta e^{i\phi_1} \spa x^3 + i x^4 = \sin \theta
e^{i\phi}
\end{equation}
where $0 \leq \theta \leq \pi/2$. We then parameterize the domain as
\begin{equation}
z^1 = \sin \theta \cos \phi \spa z^2 = \sin \theta \sin \phi
\end{equation}
We see that the $\phi_1$ circle shrinks to zero at the boundary of
the disc corresponding to $\theta= \pi/2$ while in the center of the
disc the $\phi$ circle shrinks to zero.

 {\bf Two-sphere topology:}
We consider here a domain with the shape of a disc $(z^1-z_0^1)^2 + (z^2-z^2_0 )^2 \leq 1$. Write the embedding of the two-sphere as
\begin{equation}
x^1 + i x^2 = \cos \theta e^{i\phi} \spa x^3 = \sin \theta
\end{equation}
where $0 \leq \theta \leq \pi$. We then parameterize the domain as
\begin{equation}
z^1 = z_0^1 + \cos \theta \cos \phi \spa z^2 = z_0^2 + \cos \theta \sin \phi
\end{equation}
This domain has two sheets: One sheet corresponding to $0 \leq \theta \leq \pi/2$ ($i.e.$ when $x^3 \geq 0$) and the other corresponding to $\pi/2 < \theta \leq \pi$ ($i.e.$ when $x^3 < 0$).

 {\bf Five-sphere topology:}
We consider here three circles parameterized by $\phi_1$, $\phi_2$
and $\phi_3$ fibred over a filled triangle $\frac{3}{2}|z^1| - 1
\leq z^2 \leq \frac{1}{2}$. Write the embedding of the five-sphere
as
\begin{equation}
x^1 + i x^2 = \sin \theta e^{i\phi_1} \spa x^3 + i x^4 = \cos \theta
\sin \psi e^{i\phi_2} \spa x^5 + i x^6 = \cos \theta \cos \psi
e^{i\phi_3}
\end{equation}
where $0 \leq \theta, \psi \leq \pi/2$. We then parameterize the
domain as
\begin{equation}
z^1 = \cos^2 \theta \cos 2\psi \spa z^2 = \frac{3}{2} \cos^2 \theta
-1
\end{equation}
We see that the $\phi_1$ circle shrinks to zero at the side of the
triangle with $z^2 = \frac{1}{2}$. Instead the $\phi_2$ circle
shrinks to zero at the side with $z^2 = \frac{3}{2}z^1 -1$ while the
$\phi_3$ circle shrinks to zero at the side with $z^2 =
-\frac{3}{2}z^1 - 1$.

\end{appendix}

\small


\providecommand{\href}[2]{#2}\begingroup\raggedright\endgroup

\end{document}